\documentclass[10pt,aip,jcp,superscriptaddress,twocolumn,floatfix]{revtex4-2}

\usepackage{graphicx}
\usepackage{xcolor}
\usepackage{url} 
\usepackage{hyperref} 
\usepackage{dcolumn}
\usepackage{bm}
\usepackage{multirow}
\usepackage[symbol]{footmisc}
\usepackage{amsmath}
\usepackage{amssymb}
\usepackage{amsfonts}
\usepackage{subfigure}
\usepackage{url}
\usepackage{soul}

\begin{document}

\title{Mesoscale self-organization of polydisperse magnetic nanoparticles at the water surface}

\author{Victor Ukleev}
\email{victor.ukleev@helmholtz-berlin.de}
\affiliation{Helmholtz-Zentrum Berlin f\"ur Materialien und Energie, D-14109 Berlin, Germany}

\author{Artoem Khassanov}
\affiliation{Institute of Polymer Materials of the Department of Materials Science Friedrich-Alexander University Erlangen-N\"{u}rnberg Martensstrasse 7, D-91058 Erlangen }
\affiliation{European Synchrotron Radiation Facility, 71, Avenue des Martyrs, CS40220, F-38043 Grenoble Cedex 9, France}

\author{Irina Snigireva}
\affiliation{European Synchrotron Radiation Facility, 71, Avenue des Martyrs, CS40220, F-38043 Grenoble Cedex 9, France}

\author{Oleg Konovalov}
\affiliation{European Synchrotron Radiation Facility, 71, Avenue des Martyrs, CS40220, F-38043 Grenoble Cedex 9, France}

\author{Alexei Vorobiev}
\email{avorobiev@ill.fr}
\affiliation{European Synchrotron Radiation Facility, 71, Avenue des Martyrs, CS40220, F-38043 Grenoble Cedex 9, France}
\affiliation{Department of Physics and Astronomy, Uppsala University, Box 516, 751 20, Uppsala, Sweden}


\date{\today}

\begin{abstract}
In this study, we investigated the self-ordering process in Langmuir films of polydisperse iron oxide nanoparticles on a water surface, employing in-situ X-ray scattering, surface pressure-area isotherm analysis, and Brewster angle microscopy. X-ray reflectometry confirmed the formation of a monolayer, while grazing incidence small-angle X-ray scattering revealed short-range lateral correlations with a characteristic length equal to the mean particle size. Remarkably, our findings indicated that at zero surface pressure, the particles organized into submicrometer clusters, merging upon compression to form a homogeneous layer. These layers were subsequently transferred to a solid substrate using the Langmuir-Schaefer technique and further characterized via scanning electron microscopy and polarized neutron reflectometry. Notably, our measurements unveiled a second characteristic length in the lateral correlations, orders of magnitude longer than the mean particle diameter, with polydisperse particles forming circular clusters densely packed in a hexagonal lattice. Furthermore, our evidence suggested that the lattice constant of this mesocrystal depended on the characteristics of the particle size distribution, specifically the mean particle size and the width of the size distribution. Additionally, we observed internal size separation within these clusters, where larger particles were positioned closer to the center of the cluster. Finally, polarized neutron reflectometry measurements provided valuable insights into the magnetization profile across the layer.

\end{abstract}

\maketitle

\section{Introduction}

Ordered arrays of magnetic nanoparticles (MNPs) offer promising opportunities for innovative material design, allowing for the fine-tuning of both individual MNP properties (such as size, magnetic moment, volume, and anisotropy) and the interactions between MNPs. These opportunities can be harnessed to enhance the density of bit-patterned magnetic recording media  \cite{sun2000monodisperse, kinge2008self}, develop novel sensors and optoelectronic devices  \cite{parviz2003using, courty2007large, yang2006rational}, and design catalyst surfaces \cite{long2011synthesis}. Furthermore, the collective behavior of nanoparticle assemblies holds significant relevance for the fundamental study of self-ordering phenomena.

Various methods can be employed to create ordered single-crystalline monolayers of MNPs, and the choice of synthesis conditions can significantly influence the outcome. In the past decade, it has been demonstrated  \cite{guo2003patterned, liu2009magnetic, wen2011ultra, pauly2011monolayer, pauly2009large} that highly ordered, large-area arrays of MNPs can be assembled on the surface of water using the Langmuir technique. The interactions among nanoparticles on the water surface are primarily governed by magnetic dipole-dipole forces, steric effects, and Van der Waals forces. These interactions are intricately linked to the properties of the surfactant shell, nanoparticle volume, and interparticle distance. Recent studies have shown that the ordering of monodisperse MNPs is strongly dependent on their size and magnetic moment. For instance, 10 nm iron oxide particles readily form large, highly ordered monolayers, whereas 15 nm and 20 nm MNPs, as well as binary mixtures of 10 nm and 20 nm MNPs, fail to assemble into stable monolayers and instead form three-dimensional structures \cite{vorobiev2015substantial,stanley2015spontaneous,ukleev2017self}.Furthermore, long-range ordering can be observed in some multicomponent instances \cite{yang2023self,marino2023crystallization}, even within polydisperse two-dimensional (2D) and three-dimensional (3D) nanoparticle assemblies \cite{ohara1995crystallization, xia2011self, soulantica2003spontaneous,rabideau2007}. Consequently, the self-organization of initially disordered systems, such as polydisperse ferrofluids, through diverse pathways holds the potential to facilitate the controlled synthesis of mesocrystals for a wide range of materials  \cite{sturm2017mesocrystals,jehannin2019new}.

Bulk-sensitive small-angle scattering is a valuable tool for investigating the bulk structural and magnetic properties of mesoscopic systems \cite{li2016small, honecker2022using}. Conversely, surface X-ray and neutron scattering techniques are well-suited for both in-situ and ex-situ characterization of Langmuir monolayers \cite{narayanan2017recent}. By employing a combination of X-ray reflectometry (XRR), grazing-incidence small-angle X-ray scattering (GISAXS), and spin-polarized neutron reflectometry (PNR), it becomes feasible to explore the electronic, nuclear, and magnetic structure and dynamics of these intricate systems in three dimensions \cite{narayanan2017recent, narayanan2020synchrotron}.

In our current study, we leveraged the Langmuir technique to fabricate arrays of polydisperse iron oxide MNPs on the liquid surface of water. These assemblies were subjected to in-situ surface X-ray scattering analysis. Specifically, a specular XRR experiment at the air/liquid interface was conducted to investigate the Langmuir film profile in the out-of-plane direction. Additionally, GISAXS was employed to examine in-plane interparticle correlations. We complemented these structural measurements with surface pressure-area isotherm recording and Brewster angle microscopy (BAM). Furthermore, ex-situ scanning electron microscopy (SEM) was carried out to inspect the resulting film after its deposition on a solid substrate by means of Langmuir-Schaefer technique. To gain insights into the depth distribution of the magnetic moment within the thin film post-deposition, we conducted a PNR experiment.

\section{Experimental}

\begin{figure}[]
\centering
\includegraphics[width=0.4\textwidth]{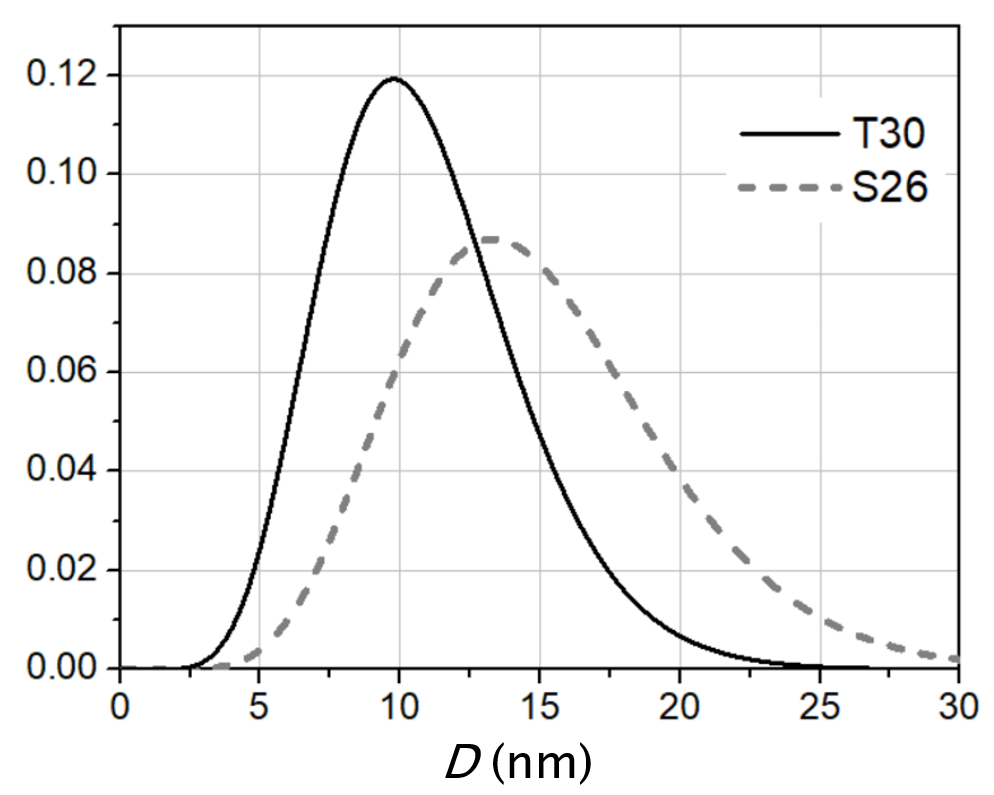}\\
\caption{The distribution of the particle size $D$ in the samples IO-T30 and IO-S26 (curves are normalized to have area equal to 1)
}
\label{size}
\end{figure}

\begin{figure}[]
\centering
\includegraphics[width=0.49\textwidth]{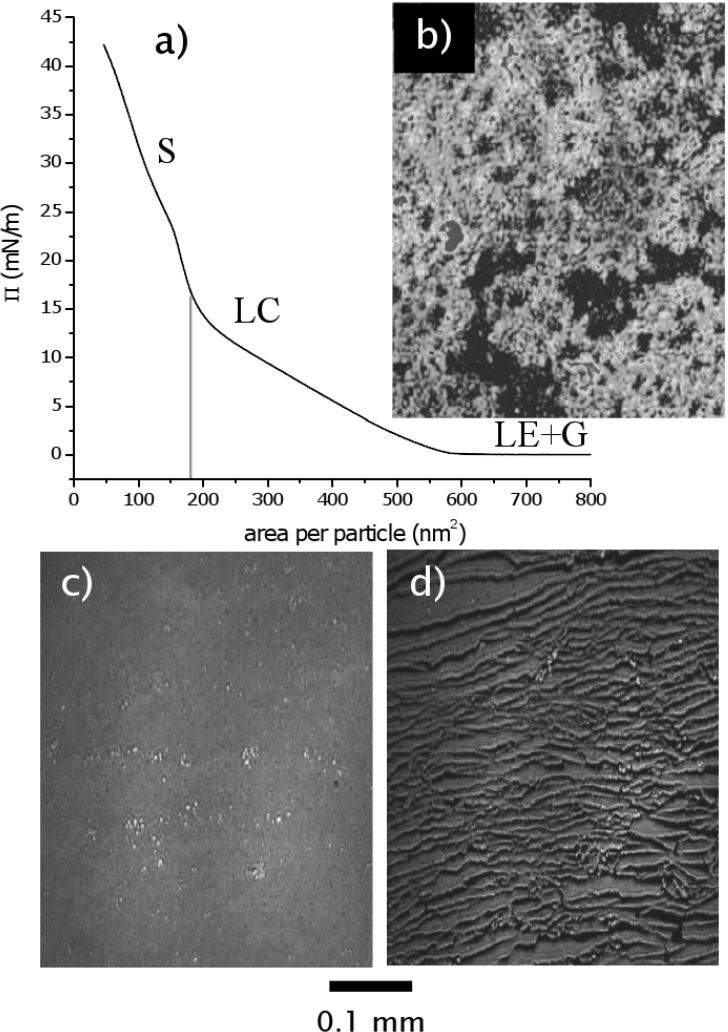}\\
\caption{(a) Surface pressure isotherms obtained for polydisperse IO-T30 particles spread on water. Grey vertical line indicates value of a nominal area occupied by a particle of radius $r_p=5$~nm covered with a surfactant layer of thickness $l_s=2.5$~nm. BAM images at (b) 1\,mN/m, (c) film relaxed to 20\,mN/m after original compression to 35\,mN/m, (d) after decompression to 6\,mN/m. The bright areas correspond to the particles, and the dark areas correspond to the water surface.}
\label{T32BAM}
\end{figure}

The iron oxide (IO) nanoparticles used in this study were synthesized through the chemical deposition of dispersed magnetite \cite{berkovsky1993magnetic}. The fabricated MNPs were coated with a layer of sodium oleate, selected by weight, and dispersed in chloroform. Information about the size distribution was obtained from wide-angle synchrotron diffraction and transmission electron microscopy (TEM) \cite{vorobiev2008nondestructive}. The size distribution followed a gamma distribution with a mean size value of $\overline{D}=10.8$ nm and a dispersion of $\sigma=3.5$ nm, achieved after subsequent centrifugation. This sample is denoted as IO-T30.

The Langmuir films were prepared using a custom-designed Langmuir trough directly installed on the sample goniometer, with the support of an active antivibration device, Halcyonics MOD2-S. The maximum and minimum subphase surface areas were 456 cm² and 115 cm², respectively. After spreading the nanoparticles, the solvent was allowed to evaporate for 15 minutes before measurement. The surface pressure was continuously monitored and recorded during the Langmuir film formation using a Wilhelmy plate made of Whatman paper and a microbalance, Model PS4 from Nima Technology Ltd.

For ex-situ observation of the monolayers forming directly on the water surface, a Brewster Angle Microscope equipped with a CCD camera and $10\times$ and $20\times$ lenses was employed, along with a separate Langmuir trough.

SEM measurements of the MNP films transferred onto solid substrates via the Langmuir-Schaefer method were conducted using a LEO 1530 microscope at ESRF.

\begin{figure}[]
\centering
\includegraphics[width=0.45\textwidth]{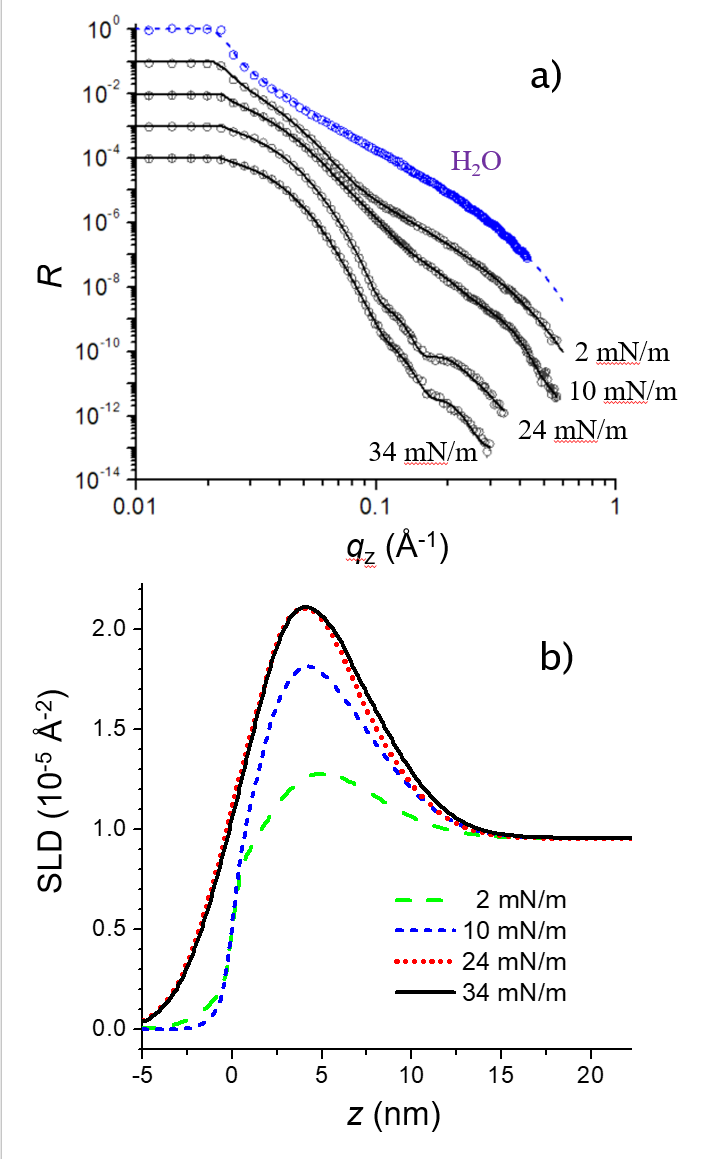}\\
\caption{(a) Experimental XRR data for IO-T30 particles on water surface (gray symbols) together with the best fit model curves (black lines) together with corresponding reference data for water (blue symbols and curve). (b) Best fit model SLD profiles as obtained at different pressure values. }
\label{T32XRR}
\end{figure}

\begin{figure}[]
\centering
\includegraphics[width=0.45\textwidth]{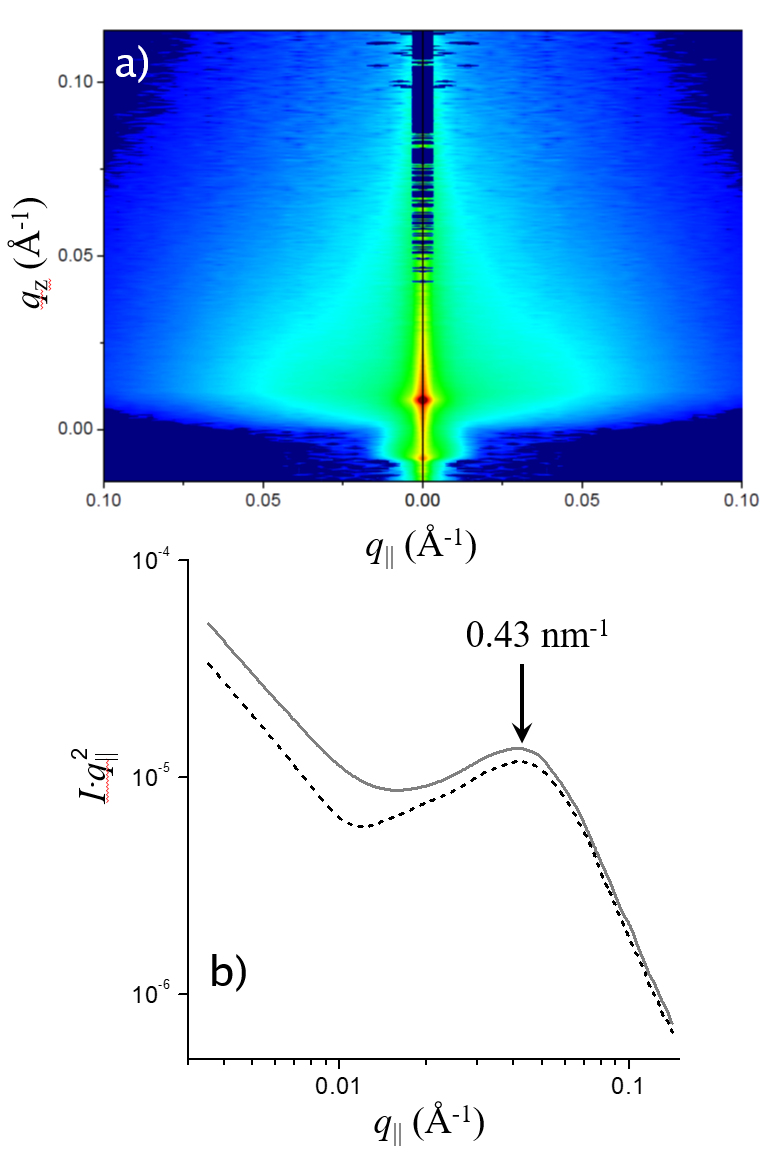}\\
\caption{(a) Two-dimensional GISAXS pattern obtained from IO-T30 sample on water surface at $\Pi = 34$\,mN/m (a). (b) Evolution of the GISAXS intensity along the cut taken at $q_z=q_c$ in the course of compression: dashed line 10\,mN/m, solid line 34\,mN/m (d). Note log-log scale. }
\label{T30GISAXS}
\end{figure}

\begin{figure*}
\centering
\includegraphics[width=0.99\textwidth]{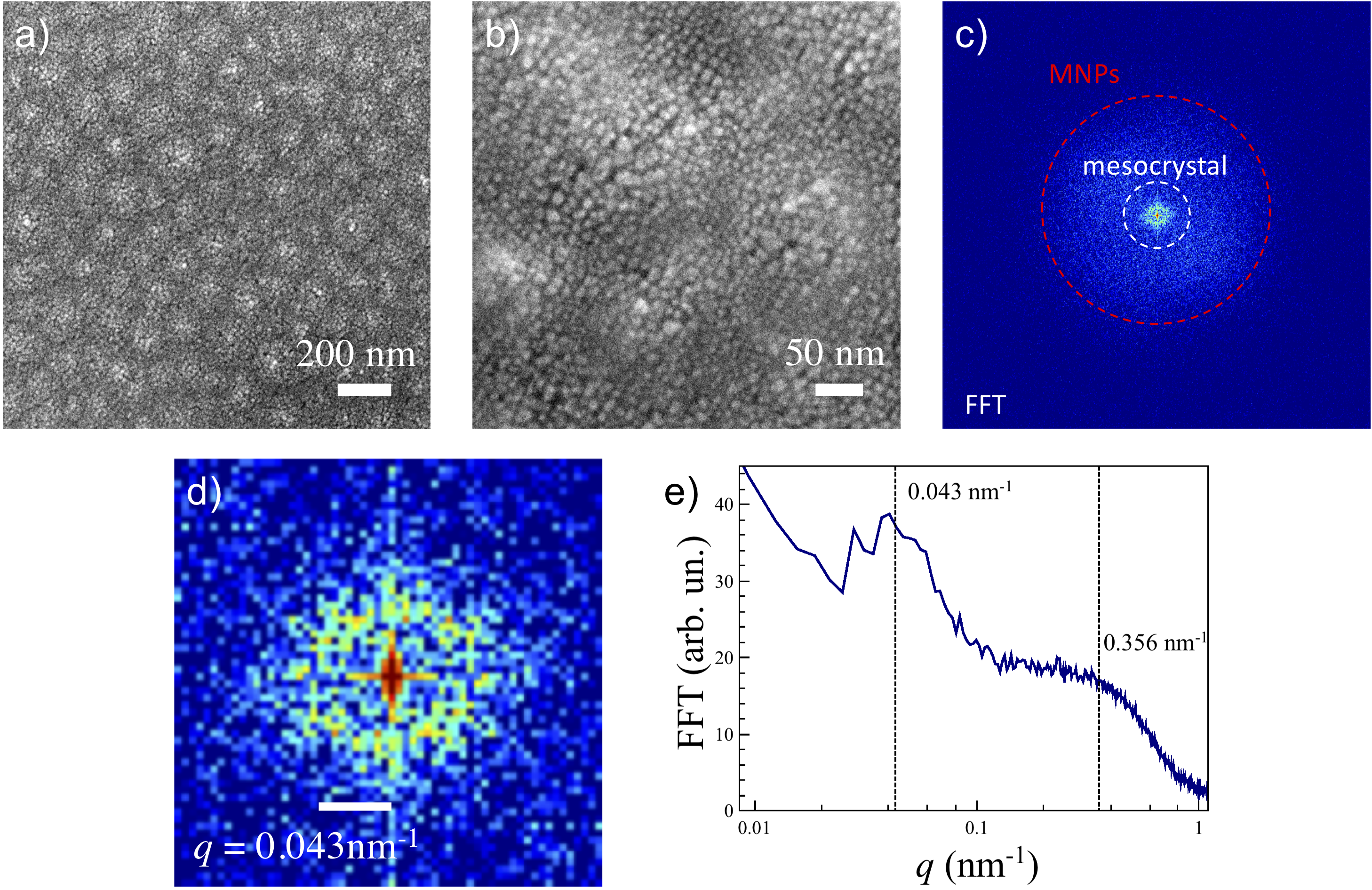}\\
\caption{(a,b) SEM images of IO-T30 layer with different magnifications. (c) Corresponding FFT image showing two contributions: individual polydisperse MNPs and orientationally-disordered mesocrystal of the nanoparticle clusters. (d) Magnification of the low-$q$ part of the FFT pattern. (e) Radially averaged FFT showing two characteristic correlation lengths.}
\label{T32SEM}
\end{figure*}

The in-situ XRR and GISAXS measurements on the liquid surface were conducted at the ID10B beamline at ESRF in Grenoble, France. The beam size was $300 \times 100$ $\mu$m$^2$ (horizontal $\times$ vertical), and the beam energy was 8 keV, corresponding to a wavelength of $\lambda = 1.54$~\AA. Two-dimensional PILATUS 300K and linear Vantec detectors were utilized for the GISAXS and XRR measurements, respectively.

Ex-situ PNR experiments on the IO-T30 sample, which had been transferred onto a Si substrate, were performed using the neutron reflectometer Super ADAM at the Institut Laue-Langevin in Grenoble, France \cite{vorobiev2015recent}. A monochromatic neutron beam with a wavelength of $\lambda=5.2$\,\AA~and the incoming polarization of $P_0=99.8\%$ was employed. The intensity of the scattered neutron beam was detected using a two-dimensional $^3$He detector DENEX 300TN. A magnetic field of $H=7$ kOe, applied in the sample plane, was generated by an electromagnet.

\section{Results and discussion}

Polydisperse IO-T30 nanoparticles were deposited in the Langmuir trough for film formation. The pressure-area isotherm for this sample, recorded during the compression process, is presented in Figure \ref{T32BAM}. Following the Harkins notation \cite{20}, this curve exhibits three distinct regions: the liquid expanded and gaseous state (LE+G) at pressures $\Pi = 0$ to 12 mN/m, the liquid condensed state (LC) at $\Pi = 13$ to 25 mN/m, and the solid state (S) beyond 25 mN/m. The LC to S transition occurs when the available area for one particle reaches the calculated value of the area occupied by a particle with the average size of 10\,nm.

BAM image (Fig. \ref{T32BAM}b) depict the formation of solid clusters immediately after deposition (Fig. \ref{T32BAM}b), corresponding to the LE+G state. Subsequently, a homogeneous coverage of the water surface is observed in the relaxed ensemble at $\Pi = 20$ mN/m, following an initial compression to 35 mN/m (LC) (Fig. \ref{T32BAM}c). Further relaxation leads to the formation of the hashed structure at $\Pi = 5$ mN/m (LE) state, as shown in Figure \ref{T32BAM}d.

Significantly, the surface pressure isotherm exhibited by the polydisperse sample can be viewed as a superposition, containing the distinctive features observed in the isotherms of monodisperse nanoparticles as reported in Ref. \cite{vorobiev2015substantial}. Consequently, the S regime corresponds to small particles with an approximate size of 10 nm, the LC regime aligns with medium-sized particles of around 15 nm, and the LE+G regime corresponds to larger particles with an approximate size of 20 nm. This comparative analysis provides valuable insights into the surface behavior of polydisperse mixtures, revealing how different size fractions contribute to the overall self-organization dynamics, drawing parallels with the behavior observed in monodisperse systems.

\begin{figure*}
\centering
\includegraphics[width=0.99\textwidth]{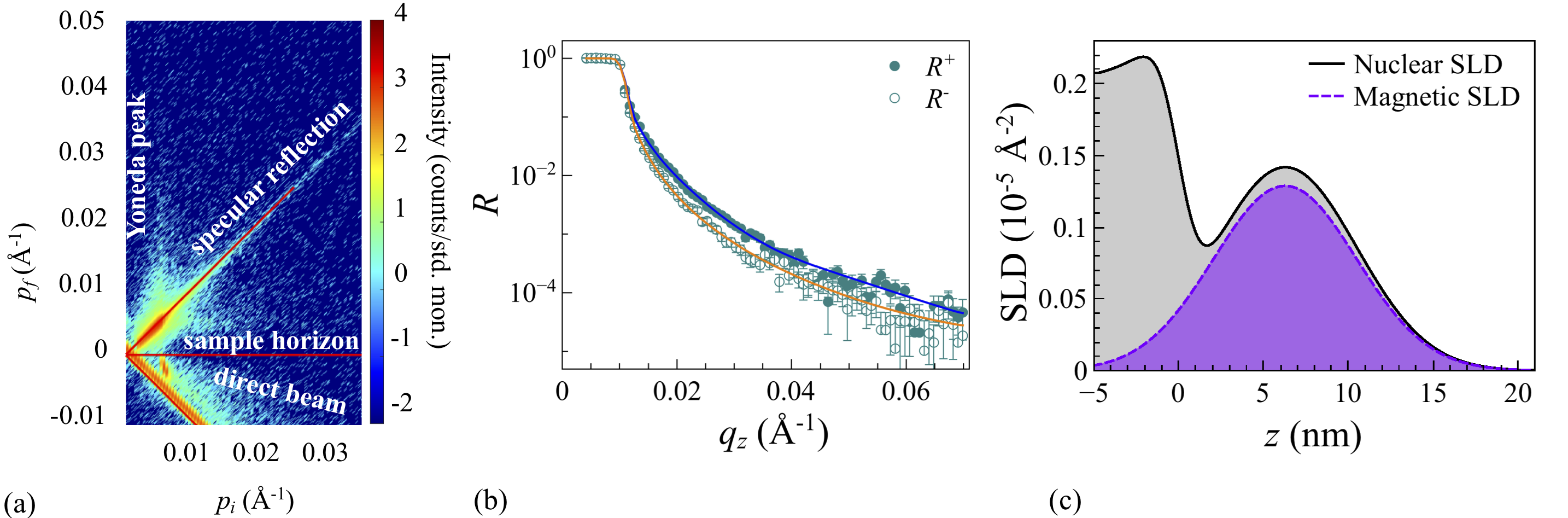}\\
\caption{Two-dimensional experimental PNR intensity for IO-T30 particles deposited onto Si wafer as a function of the projection of the incoming and outgoing wavevectors normal to the sample plane ($p_i = (2 \pi/\lambda) \sin{\alpha_i}$ and $p_f = (2 \pi/\lambda) \sin{\alpha_f}$, where $\alpha_i$ and $\alpha_f$ are incident and scattering angles, respectively) (a). Integrated specular reflectivity intensities measured with neutron polarization parallel ($R^+$) and antiparallel ($R^-$) to the applied magnetic field $H=7$\,kOe as a function of momentum transfer vector $q_z$ (b). Solid lines represents the fitted curves according to the reconstructed nuclear and magnetic SLD profiles of the film shown in panel (c).}
\label{T32PNR}
\end{figure*}

The XRR curves, along with the corresponding electron scattering length density (SLD) profiles for the IO-T30 sample, obtained from the fit, are presented in Figure \ref{T32XRR}. As an example, the experimental XRR curve from the pure air/water interface, fitted with the Fresnel decay function, is also included. Due to the particle size distribution in the polydisperse sample, the roughness of the layer is of the same order of magnitude as the layer thickness. As a result, no thickness oscillations are observed in the XRR curve at the low coverage regime (LE+G), owing to the rapid decay of the reflected intensity induced by the roughness. The first Kissieg fringe is evident in the XRR curve at pressure values of $\Pi = 24$ mN/m and 34 mN/m when the layer reaches the S state and becomes more homogeneous. The model incorporates a layer with a wide distribution of electron density, corresponding to the size distribution of the nanoparticles (Fig. \ref{T32XRR}b). Further compression of the layer leads to an increase in the SLD, indicating the elimination of gaps between the nanoparticles, while the total film thickness remains constant, and no transition to a bi- or multilayered structure (hashing) is observed.

In-plane correlations of the polydisperse IO-T30 sample were studied using GISAXS (Fig. \ref{T32XRR}c) at a constant incident angle $\alpha_i = 0.13^\circ$, which is slightly below the critical angle $\alpha_c = 0.15^\circ$, at two surface pressures of $\Pi = 10$ mN/m and $\Pi = 44$ mN/m. The scattered intensity exhibits a nearly homogeneous decay along the $q_{||}$ vector without the appearance of Bragg peaks. Images at both pressures suggest the absence of inter-particle correlations in the Langmuir film, primarily due to the high polydispersity of the nanoparticles. To determine the mean in-plane interparticle distance parameter $a$, the Krattky representation $I\cdot q_{||}^2$ of the intensity along the cut taken at $q_z = q_c$ was utilized (Fig. \ref{T32XRR}d). The mean in-plane interparticle distance was found to be $a = 14.9(1)$ nm, with an average coherent domain size of $D = 230$ nm.

Figures \ref{T32SEM}a and b depict the IO-T30 sample deposited on a gold thiol-treated surface from the Langmuir layer at $\Pi=34$ mN/m. The SEM images reveal a non-uniform, disordered layer of three-dimensional clusters of nanoparticles with varying sizes. The large-scale image (Fig. \ref{T32SEM}b) illustrates that larger MNPs aggregate into clusters with an average size of 200 nm, surrounded by smaller particle domains. This size separation is likely driven by magnetic dipolar interactions, favoring the agglomeration of MNPs with larger volume and, consequently, greater magnetic moment. This magnetically-induced self-separation mechanism has been previously proposed for simplified systems consisting of binary assemblies of monodisperse nanoparticles \cite{stanley2015spontaneous,ukleev2017self}. Interestingly, the clusters form a short-range ordered mesocrystal on the scale of hundreds of nanometers, visible in the real-space image (Fig. \ref{T32SEM}a) but not in the GISAXS pattern, as the instrument configuration was tuned to detect signals from smaller objects at higher scattering angles. The mean interparticle distance $a=17.7$ nm, derived from the radial average of the fast Fourier transformed (FFT) SEM image (Fig. \ref{T32SEM}c), is in reasonable agreement with the in-situ experiment results, accounting for the local nature of the SEM probe.

The manifestation of large ($>100$ nm) ferromagnetic domains in the ordered nanostructures of iron oxide nanoparticles has been recently revealed through neutron scattering \cite{mishra2012self, theis2018self, bender2018dipolar,theis2020self}, resonant soft X-ray scattering \cite{chesnel2018unraveling}, and off-axis electron holography \cite{wang2019manipulation}. In this study, PNR was applied to investigate the out-of-plane distribution of magnetization within the nanoparticle film after deposition onto a solid substrate. Previous reports have unequivocally demonstrated the effectiveness of this approach, even in the case of very thin films, such as monolayers of MNPs with a 10 nm diameter \cite{ukleev2017polarized}. Theoretical descriptions and experimental details can be found in Ref. \onlinecite{zhu2005modern}. The two-dimensional PNR $(p_i, p_f)$ map obtained from the IO-T30 sample is presented in Figure \ref{T32PNR}a, featuring the specular and off-specular contributions of neutron reflection from the film. The integrated specular reflection curves $R^{+(-)}(q_z)$, measured with opposite neutron polarizations in the saturating magnetic field $H=7$ kOe applied in the sample plane, were fitted using the Parratt algorithm within the GenX software package \cite{bjorck2007genx,glavic2022genx} (Fig. \ref{T32PNR}b).The magnitude of the applied field is sufficient to fully saturate the magnetization in the sample plane \cite{pichon2014magnetic}. The PNR curves were fitted with the same model as the XRR. The model allowed a reduced nuclear SLD $\rho_n$ compared to the in-situ X-ray study in order to account for possible density reduction of the film during the transfer from the water subphase to the solid substrate \cite{ukleev2016x}. The magnetic SLD part $\rho_m$ was allowed to vary freely. 

As a result of the fitting routine, nuclear and magnetic SLD profiles were obtained, as shown in Figure \ref{T32PNR}c. Similar to the electron density profiles obtained in the XRR study, the nuclear and magnetic densities of the film are primarily concentrated in an approximately 15 nm-thick layer above the substrate ($z = 0$). The round-shaped profile with a maximum at $z\approx6.3$ nm from the Si surface corresponds to the center of the MNPs layer. The magnetic SLD is directly proportional to the in-plane net magnetization component $M\textrm{[kA/m]}=3.5\cdot10^9\rho_m\textrm{[\AA$^{-2}$]}$ \cite{zhu2005modern}. According to this equation, the peak value of magnetization is $M=450\pm24$ emu/cm$^3$, matching the expected value for magnetite (480 kA/m$^3$) \cite{barbeta2010magnetic}. Therefore, the shape of the neutron SLD confirms the chemical depth profile obtained from XRR and the planar homogeneity of the film observed by SEM. The magnetic depth profile reflects the original particle size distribution in the ferrofluid, with magnetization values enhanced compared to the monodisperse case of IO-10 \cite{ukleev2017polarized}. The absence of any notable peaks in the off-specular PNR signal (Fig. \ref{T32PNR}a), which is sensitive to lateral correlations with sizes from a few hundred nm to a few microns, implies that there is no long-range order formed by the mesoscale clusters.

\begin{figure}
\centering
\includegraphics[width=0.4\textwidth]{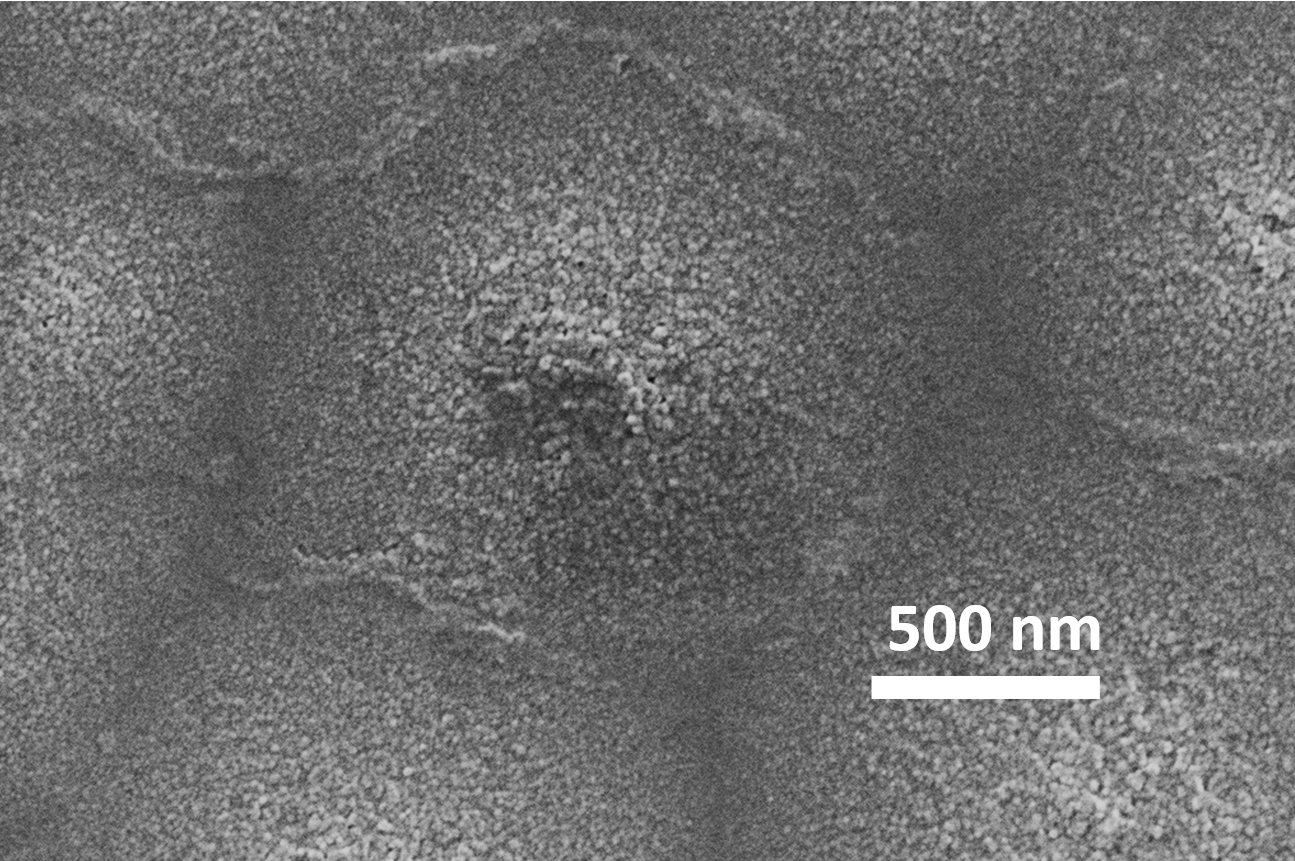}\\
\caption{SEM image of IO-S26 sample showing a large cluster of polydisperse MNPs.}
\label{S26SEM}
\end{figure}

Previous studies have shown that formation of mixed clusters of nanoparticles is  energetically favourable in binary mixtures of monodisperse MPNs as a result of the interplay between magnetic dipolar, van der Waals interactions, and steric repulsion \cite{stanley2015spontaneous,ukleev2017self}. Similar mechanisms should also apply to the polydisperse mixtures with comparable mean particle sizes. Notably, within the polydisperse ensemble featuring slightly larger average particle sizes, such as in the case of S26 (Fig. \ref{size}a), the phenomenon of size-separated mesoscale structures becomes even more pronounced. As illustrated in Fig. \ref{S26SEM}, the self-organization of micrometer-scale clusters of MNPs is evident, with larger particles concentrated in the central region and smaller ones at the periphery. We hypothesize that the increased magnetic moment of this assembly plays a crucial role in fostering the formation of these more substantial clusters, distinguishing it from the behavior observed in the T30 configuration. This observation underscores the significant influence of both particle size and magnetic characteristics on the intricate self-assembly processes within polydisperse nanoparticle ensembles.

Recent studies have unveiled that small monodisperse self-assembled particles with diameters less than or equal to 20 nm tend to exhibit not only disordered (superparamagnetic) behavior but also anti-ferromagnetic ordering \cite{frandsen2021superparamagnetic,chesnel2018unraveling, bender2018dipolar,rackham2023field}. Consequently, further exploration of mesoscopic magnetic textures that may emerge atop the structural clusters in polydisperse samples represents an intriguing avenue for future research.

\section{Conclusion}

The observations made in this study provide valuable insights into the self-organization of polydisperse iron oxide nanoparticles on a water surface. While previous research has primarily focused on the self-ordering of monodisperse nanoparticles, the behavior of polydisperse systems adds complexity to the understanding of these processes. In contrast to the self-ordering observed in monodisperse MNPs reported in Ref. \cite{vorobiev2015substantial}, where 10 nm, 15 nm, and 20 nm MNPs form in-plane ordered structures, the polydisperse mixture IO-T30 tends to assemble into a more homogeneous amorphous monolayer. Furthermore, a short-range ordered mesocrystal of larger particle agglomerates is formed due to a magnetically-driven self-separation process, similar to what has been previously investigated in the case of binary mixtures \cite{stanley2015spontaneous,ukleev2017self}. The fact that polydisperse particles tend to form a more homogeneous amorphous monolayer suggests that size variations play a significant role in the self-assembly process. This observation has practical implications for the controlled synthesis of mesocrystals, as it highlights the need to consider particle size distributions when designing nanoparticle-based materials. Formation of the mesoscale clusters of MNPs generates another degree of freedom that can be used to control the self-assembly, for example, by aligning the clusters into long-range ordered structures using external magnetic fields \cite{sahoo2004field,tracy2013magnetic,mehdizadeh2015self}.

These findings contribute to our understanding of self-assembly processes in complex nanoparticle systems. The insights gained here can inform the design and synthesis of materials with tailored properties, opening up new possibilities for applications in areas such as magnetic data storage, sensing, and catalysis.

Further investigations into the magnetic textures that may arise within polydisperse nanoparticle assemblies promise exciting prospects for future research. Understanding the interplay between structural and magnetic properties in these systems could lead to the development of novel materials with unique functionalities.
 
\section*{Acknowledgments}

Authors thank European Synchrotron Radiation Facility and Institut Laue-Langevin for provided beamtime and technical assistance.

\section*{Author Declarations}
\subsection*{Conflict of Interest}
The authors have no conflicts to disclose.
\subsection*{Author Contributions}
V.U., A.K., I.S., O.K., A.V performed experiments, V.U., A.V. analyzed the data and wrote the manuscript, V.U., O.K., and A.V. jointly conceived the project. All authors read and edited the manuscript.

\section*{Data availability Statement}
Data is available from V.U. and A.V. upon request.  

%


\begin{thebibliography}{48}%
\makeatletter
\providecommand \@ifxundefined [1]{%
 \@ifx{#1\undefined}
}%
\providecommand \@ifnum [1]{%
 \ifnum #1\expandafter \@firstoftwo
 \else \expandafter \@secondoftwo
 \fi
}%
\providecommand \@ifx [1]{%
 \ifx #1\expandafter \@firstoftwo
 \else \expandafter \@secondoftwo
 \fi
}%
\providecommand \natexlab [1]{#1}%
\providecommand \enquote  [1]{``#1''}%
\providecommand \bibnamefont  [1]{#1}%
\providecommand \bibfnamefont [1]{#1}%
\providecommand \citenamefont [1]{#1}%
\providecommand \href@noop [0]{\@secondoftwo}%
\providecommand \href [0]{\begingroup \@sanitize@url \@href}%
\providecommand \@href[1]{\@@startlink{#1}\@@href}%
\providecommand \@@href[1]{\endgroup#1\@@endlink}%
\providecommand \@sanitize@url [0]{\catcode `\\12\catcode `\$12\catcode `\&12\catcode `\#12\catcode `\^12\catcode `\_12\catcode `\%12\relax}%
\providecommand \@@startlink[1]{}%
\providecommand \@@endlink[0]{}%
\providecommand \url  [0]{\begingroup\@sanitize@url \@url }%
\providecommand \@url [1]{\endgroup\@href {#1}{\urlprefix }}%
\providecommand \urlprefix  [0]{URL }%
\providecommand \Eprint [0]{\href }%
\providecommand \doibase [0]{https://doi.org/}%
\providecommand \selectlanguage [0]{\@gobble}%
\providecommand \bibinfo  [0]{\@secondoftwo}%
\providecommand \bibfield  [0]{\@secondoftwo}%
\providecommand \translation [1]{[#1]}%
\providecommand \BibitemOpen [0]{}%
\providecommand \bibitemStop [0]{}%
\providecommand \bibitemNoStop [0]{.\EOS\space}%
\providecommand \EOS [0]{\spacefactor3000\relax}%
\providecommand \BibitemShut  [1]{\csname bibitem#1\endcsname}%
\let\auto@bib@innerbib\@empty
\bibitem [{\citenamefont {Sun}\ \emph {et~al.}(2000)\citenamefont {Sun}, \citenamefont {Murray}, \citenamefont {Weller}, \citenamefont {Folks},\ and\ \citenamefont {Moser}}]{sun2000monodisperse}%
  \BibitemOpen
  \bibfield  {author} {\bibinfo {author} {\bibfnamefont {S.}~\bibnamefont {Sun}}, \bibinfo {author} {\bibfnamefont {C.}~\bibnamefont {Murray}}, \bibinfo {author} {\bibfnamefont {D.}~\bibnamefont {Weller}}, \bibinfo {author} {\bibfnamefont {L.}~\bibnamefont {Folks}},\ and\ \bibinfo {author} {\bibfnamefont {A.}~\bibnamefont {Moser}},\ }\bibfield  {title} {\enquote {\bibinfo {title} {Monodisperse {FePt} nanoparticles and ferromagnetic {FePt} nanocrystal superlattices},}\ }\href@noop {} {\bibfield  {journal} {\bibinfo  {journal} {Science}\ }\textbf {\bibinfo {volume} {287}},\ \bibinfo {pages} {1989--1992} (\bibinfo {year} {2000})}\BibitemShut {NoStop}%
\bibitem [{\citenamefont {Kinge}, \citenamefont {Crego-Calama},\ and\ \citenamefont {Reinhoudt}(2008)}]{kinge2008self}%
  \BibitemOpen
  \bibfield  {author} {\bibinfo {author} {\bibfnamefont {S.}~\bibnamefont {Kinge}}, \bibinfo {author} {\bibfnamefont {M.}~\bibnamefont {Crego-Calama}},\ and\ \bibinfo {author} {\bibfnamefont {D.~N.}\ \bibnamefont {Reinhoudt}},\ }\bibfield  {title} {\enquote {\bibinfo {title} {Self-assembling nanoparticles at surfaces and interfaces},}\ }\href@noop {} {\bibfield  {journal} {\bibinfo  {journal} {ChemPhysChem}\ }\textbf {\bibinfo {volume} {9}},\ \bibinfo {pages} {20--42} (\bibinfo {year} {2008})}\BibitemShut {NoStop}%
\bibitem [{\citenamefont {Parviz}, \citenamefont {Ryan},\ and\ \citenamefont {Whitesides}(2003)}]{parviz2003using}%
  \BibitemOpen
  \bibfield  {author} {\bibinfo {author} {\bibfnamefont {B.~A.}\ \bibnamefont {Parviz}}, \bibinfo {author} {\bibfnamefont {D.}~\bibnamefont {Ryan}},\ and\ \bibinfo {author} {\bibfnamefont {G.~M.}\ \bibnamefont {Whitesides}},\ }\bibfield  {title} {\enquote {\bibinfo {title} {Using self-assembly for the fabrication of nano-scale electronic and photonic devices},}\ }\href@noop {} {\bibfield  {journal} {\bibinfo  {journal} {Advanced Packaging, IEEE Transactions on}\ }\textbf {\bibinfo {volume} {26}},\ \bibinfo {pages} {233--241} (\bibinfo {year} {2003})}\BibitemShut {NoStop}%
\bibitem [{\citenamefont {Courty}\ \emph {et~al.}(2007)\citenamefont {Courty}, \citenamefont {Henry}, \citenamefont {Goubet},\ and\ \citenamefont {Pileni}}]{courty2007large}%
  \BibitemOpen
  \bibfield  {author} {\bibinfo {author} {\bibfnamefont {A.}~\bibnamefont {Courty}}, \bibinfo {author} {\bibfnamefont {A.-I.}\ \bibnamefont {Henry}}, \bibinfo {author} {\bibfnamefont {N.}~\bibnamefont {Goubet}},\ and\ \bibinfo {author} {\bibfnamefont {M.-P.}\ \bibnamefont {Pileni}},\ }\bibfield  {title} {\enquote {\bibinfo {title} {Large triangular single crystals formed by mild annealing of self-organized silver nanocrystals},}\ }\href@noop {} {\bibfield  {journal} {\bibinfo  {journal} {Nature Materials}\ }\textbf {\bibinfo {volume} {6}},\ \bibinfo {pages} {900--907} (\bibinfo {year} {2007})}\BibitemShut {NoStop}%
\bibitem [{\citenamefont {Yang}\ \emph {et~al.}(2006)\citenamefont {Yang}, \citenamefont {Elim}, \citenamefont {Zhang}, \citenamefont {Lee},\ and\ \citenamefont {Ji}}]{yang2006rational}%
  \BibitemOpen
  \bibfield  {author} {\bibinfo {author} {\bibfnamefont {J.}~\bibnamefont {Yang}}, \bibinfo {author} {\bibfnamefont {H.~I.}\ \bibnamefont {Elim}}, \bibinfo {author} {\bibfnamefont {Q.}~\bibnamefont {Zhang}}, \bibinfo {author} {\bibfnamefont {J.~Y.}\ \bibnamefont {Lee}},\ and\ \bibinfo {author} {\bibfnamefont {W.}~\bibnamefont {Ji}},\ }\bibfield  {title} {\enquote {\bibinfo {title} {Rational synthesis, self-assembly, and optical properties of {PbS-Au} heterogeneous nanostructures via preferential deposition},}\ }\href@noop {} {\bibfield  {journal} {\bibinfo  {journal} {Journal of the American Chemical Society}\ }\textbf {\bibinfo {volume} {128}},\ \bibinfo {pages} {11921--11926} (\bibinfo {year} {2006})}\BibitemShut {NoStop}%
\bibitem [{\citenamefont {Long}\ \emph {et~al.}(2011)\citenamefont {Long}, \citenamefont {Ohtaki}, \citenamefont {Uchida}, \citenamefont {Jalem}, \citenamefont {Hirata}, \citenamefont {Chien},\ and\ \citenamefont {Nogami}}]{long2011synthesis}%
  \BibitemOpen
  \bibfield  {author} {\bibinfo {author} {\bibfnamefont {N.~V.}\ \bibnamefont {Long}}, \bibinfo {author} {\bibfnamefont {M.}~\bibnamefont {Ohtaki}}, \bibinfo {author} {\bibfnamefont {M.}~\bibnamefont {Uchida}}, \bibinfo {author} {\bibfnamefont {R.}~\bibnamefont {Jalem}}, \bibinfo {author} {\bibfnamefont {H.}~\bibnamefont {Hirata}}, \bibinfo {author} {\bibfnamefont {N.~D.}\ \bibnamefont {Chien}},\ and\ \bibinfo {author} {\bibfnamefont {M.}~\bibnamefont {Nogami}},\ }\bibfield  {title} {\enquote {\bibinfo {title} {Synthesis and characterization of polyhedral pt nanoparticles: Their catalytic property, surface attachment, self-aggregation and assembly},}\ }\href@noop {} {\bibfield  {journal} {\bibinfo  {journal} {Journal of Colloid and Interface Science}\ }\textbf {\bibinfo {volume} {359}},\ \bibinfo {pages} {339--350} (\bibinfo {year} {2011})}\BibitemShut {NoStop}%
\bibitem [{\citenamefont {Guo}\ \emph {et~al.}(2003)\citenamefont {Guo}, \citenamefont {Teng}, \citenamefont {Rahman},\ and\ \citenamefont {Yang}}]{guo2003patterned}%
  \BibitemOpen
  \bibfield  {author} {\bibinfo {author} {\bibfnamefont {Q.}~\bibnamefont {Guo}}, \bibinfo {author} {\bibfnamefont {X.}~\bibnamefont {Teng}}, \bibinfo {author} {\bibfnamefont {S.}~\bibnamefont {Rahman}},\ and\ \bibinfo {author} {\bibfnamefont {H.}~\bibnamefont {Yang}},\ }\bibfield  {title} {\enquote {\bibinfo {title} {Patterned {Langmuir-Blodgett} films of monodisperse nanoparticles of iron oxide using soft lithography},}\ }\href@noop {} {\bibfield  {journal} {\bibinfo  {journal} {Journal of the American Chemical Society}\ }\textbf {\bibinfo {volume} {125}},\ \bibinfo {pages} {630--631} (\bibinfo {year} {2003})}\BibitemShut {NoStop}%
\bibitem [{\citenamefont {Liu}\ \emph {et~al.}(2009)\citenamefont {Liu}, \citenamefont {Shan}, \citenamefont {Zhu},\ and\ \citenamefont {Chen}}]{liu2009magnetic}%
  \BibitemOpen
  \bibfield  {author} {\bibinfo {author} {\bibfnamefont {C.}~\bibnamefont {Liu}}, \bibinfo {author} {\bibfnamefont {Y.}~\bibnamefont {Shan}}, \bibinfo {author} {\bibfnamefont {Y.}~\bibnamefont {Zhu}},\ and\ \bibinfo {author} {\bibfnamefont {K.}~\bibnamefont {Chen}},\ }\bibfield  {title} {\enquote {\bibinfo {title} {Magnetic monolayer film of oleic acid-stabilized {Fe$_3$O$_4$} particles fabricated via langmuir-blodgett technique},}\ }\href@noop {} {\bibfield  {journal} {\bibinfo  {journal} {Thin Solid Films}\ }\textbf {\bibinfo {volume} {518}},\ \bibinfo {pages} {324--327} (\bibinfo {year} {2009})}\BibitemShut {NoStop}%
\bibitem [{\citenamefont {Wen}\ and\ \citenamefont {Majetich}(2011)}]{wen2011ultra}%
  \BibitemOpen
  \bibfield  {author} {\bibinfo {author} {\bibfnamefont {T.}~\bibnamefont {Wen}}\ and\ \bibinfo {author} {\bibfnamefont {S.~A.}\ \bibnamefont {Majetich}},\ }\bibfield  {title} {\enquote {\bibinfo {title} {Ultra-large-area self-assembled monolayers of nanoparticles},}\ }\href@noop {} {\bibfield  {journal} {\bibinfo  {journal} {ACS Nano}\ }\textbf {\bibinfo {volume} {5}},\ \bibinfo {pages} {8868--8876} (\bibinfo {year} {2011})}\BibitemShut {NoStop}%
\bibitem [{\citenamefont {Pauly}\ \emph {et~al.}(2011)\citenamefont {Pauly}, \citenamefont {Pichon}, \citenamefont {Albouy}, \citenamefont {Fleutot}, \citenamefont {Leuvrey}, \citenamefont {Trassin}, \citenamefont {Gallani},\ and\ \citenamefont {Begin-Colin}}]{pauly2011monolayer}%
  \BibitemOpen
  \bibfield  {author} {\bibinfo {author} {\bibfnamefont {M.}~\bibnamefont {Pauly}}, \bibinfo {author} {\bibfnamefont {B.~P.}\ \bibnamefont {Pichon}}, \bibinfo {author} {\bibfnamefont {P.-A.}\ \bibnamefont {Albouy}}, \bibinfo {author} {\bibfnamefont {S.}~\bibnamefont {Fleutot}}, \bibinfo {author} {\bibfnamefont {C.}~\bibnamefont {Leuvrey}}, \bibinfo {author} {\bibfnamefont {M.}~\bibnamefont {Trassin}}, \bibinfo {author} {\bibfnamefont {J.-L.}\ \bibnamefont {Gallani}},\ and\ \bibinfo {author} {\bibfnamefont {S.}~\bibnamefont {Begin-Colin}},\ }\bibfield  {title} {\enquote {\bibinfo {title} {Monolayer and multilayer assemblies of spherically and cubic-shaped iron oxide nanoparticles},}\ }\href@noop {} {\bibfield  {journal} {\bibinfo  {journal} {Journal of Materials Chemistry}\ }\textbf {\bibinfo {volume} {21}},\ \bibinfo {pages} {16018--16027} (\bibinfo {year} {2011})}\BibitemShut {NoStop}%
\bibitem [{\citenamefont {Pauly}\ \emph {et~al.}(2009)\citenamefont {Pauly}, \citenamefont {Pichon}, \citenamefont {Demorti{\`e}re}, \citenamefont {Delahaye}, \citenamefont {Leuvrey}, \citenamefont {Pourroy},\ and\ \citenamefont {B{\'e}gin-Colin}}]{pauly2009large}%
  \BibitemOpen
  \bibfield  {author} {\bibinfo {author} {\bibfnamefont {M.}~\bibnamefont {Pauly}}, \bibinfo {author} {\bibfnamefont {B.~P.}\ \bibnamefont {Pichon}}, \bibinfo {author} {\bibfnamefont {A.}~\bibnamefont {Demorti{\`e}re}}, \bibinfo {author} {\bibfnamefont {J.}~\bibnamefont {Delahaye}}, \bibinfo {author} {\bibfnamefont {C.}~\bibnamefont {Leuvrey}}, \bibinfo {author} {\bibfnamefont {G.}~\bibnamefont {Pourroy}},\ and\ \bibinfo {author} {\bibfnamefont {S.}~\bibnamefont {B{\'e}gin-Colin}},\ }\bibfield  {title} {\enquote {\bibinfo {title} {Large {2D} monolayer assemblies of iron oxide nanocrystals by the {Langmuir--Blodgett} technique},}\ }\href@noop {} {\bibfield  {journal} {\bibinfo  {journal} {Superlattices and Microstructures}\ }\textbf {\bibinfo {volume} {46}},\ \bibinfo {pages} {195--204} (\bibinfo {year} {2009})}\BibitemShut {NoStop}%
\bibitem [{\citenamefont {Vorobiev}\ \emph {et~al.}(2015{\natexlab{a}})\citenamefont {Vorobiev}, \citenamefont {Khassanov}, \citenamefont {Ukleev}, \citenamefont {Snigireva},\ and\ \citenamefont {Konovalov}}]{vorobiev2015substantial}%
  \BibitemOpen
  \bibfield  {author} {\bibinfo {author} {\bibfnamefont {A.}~\bibnamefont {Vorobiev}}, \bibinfo {author} {\bibfnamefont {A.}~\bibnamefont {Khassanov}}, \bibinfo {author} {\bibfnamefont {V.}~\bibnamefont {Ukleev}}, \bibinfo {author} {\bibfnamefont {I.}~\bibnamefont {Snigireva}},\ and\ \bibinfo {author} {\bibfnamefont {O.}~\bibnamefont {Konovalov}},\ }\bibfield  {title} {\enquote {\bibinfo {title} {Substantial difference in ordering of 10, 15, and 20 nm iron oxide nanoparticles on a water surface: in situ characterization by the grazing incidence {X-ray} scattering},}\ }\href@noop {} {\bibfield  {journal} {\bibinfo  {journal} {Langmuir}\ }\textbf {\bibinfo {volume} {31}},\ \bibinfo {pages} {11639--11648} (\bibinfo {year} {2015}{\natexlab{a}})}\BibitemShut {NoStop}%
\bibitem [{\citenamefont {Stanley}\ \emph {et~al.}(2015)\citenamefont {Stanley}, \citenamefont {Boucheron}, \citenamefont {Lin}, \citenamefont {Meron},\ and\ \citenamefont {Shpyrko}}]{stanley2015spontaneous}%
  \BibitemOpen
  \bibfield  {author} {\bibinfo {author} {\bibfnamefont {J.}~\bibnamefont {Stanley}}, \bibinfo {author} {\bibfnamefont {L.}~\bibnamefont {Boucheron}}, \bibinfo {author} {\bibfnamefont {B.}~\bibnamefont {Lin}}, \bibinfo {author} {\bibfnamefont {M.}~\bibnamefont {Meron}},\ and\ \bibinfo {author} {\bibfnamefont {O.}~\bibnamefont {Shpyrko}},\ }\bibfield  {title} {\enquote {\bibinfo {title} {Spontaneous phase separation during self-assembly in bi-dispersed spherical iron oxide nanoparticle monolayers},}\ }\href@noop {} {\bibfield  {journal} {\bibinfo  {journal} {Applied Physics Letters}\ }\textbf {\bibinfo {volume} {106}},\ \bibinfo {pages} {161602} (\bibinfo {year} {2015})}\BibitemShut {NoStop}%
\bibitem [{\citenamefont {Ukleev}\ \emph {et~al.}(2017)\citenamefont {Ukleev}, \citenamefont {Khassanov}, \citenamefont {Snigireva}, \citenamefont {Konovalov}, \citenamefont {Dudnik}, \citenamefont {Dubitskiy},\ and\ \citenamefont {Vorobiev}}]{ukleev2017self}%
  \BibitemOpen
  \bibfield  {author} {\bibinfo {author} {\bibfnamefont {V.}~\bibnamefont {Ukleev}}, \bibinfo {author} {\bibfnamefont {A.}~\bibnamefont {Khassanov}}, \bibinfo {author} {\bibfnamefont {I.}~\bibnamefont {Snigireva}}, \bibinfo {author} {\bibfnamefont {O.}~\bibnamefont {Konovalov}}, \bibinfo {author} {\bibfnamefont {M.}~\bibnamefont {Dudnik}}, \bibinfo {author} {\bibfnamefont {I.}~\bibnamefont {Dubitskiy}},\ and\ \bibinfo {author} {\bibfnamefont {A.}~\bibnamefont {Vorobiev}},\ }\bibfield  {title} {\enquote {\bibinfo {title} {Self-assembly of a binary mixture of iron oxide nanoparticles in langmuir film: {X-ray} scattering study},}\ }\href@noop {} {\bibfield  {journal} {\bibinfo  {journal} {Materials Chemistry and Physics}\ }\textbf {\bibinfo {volume} {202}},\ \bibinfo {pages} {31--39} (\bibinfo {year} {2017})}\BibitemShut {NoStop}%
\bibitem [{\citenamefont {Yang}\ \emph {et~al.}(2023)\citenamefont {Yang}, \citenamefont {LaCour}, \citenamefont {Cai}, \citenamefont {Xu}, \citenamefont {Rosen}, \citenamefont {Zhang}, \citenamefont {Kagan}, \citenamefont {Glotzer},\ and\ \citenamefont {Murray}}]{yang2023self}%
  \BibitemOpen
  \bibfield  {author} {\bibinfo {author} {\bibfnamefont {S.}~\bibnamefont {Yang}}, \bibinfo {author} {\bibfnamefont {R.~A.}\ \bibnamefont {LaCour}}, \bibinfo {author} {\bibfnamefont {Y.-Y.}\ \bibnamefont {Cai}}, \bibinfo {author} {\bibfnamefont {J.}~\bibnamefont {Xu}}, \bibinfo {author} {\bibfnamefont {D.~J.}\ \bibnamefont {Rosen}}, \bibinfo {author} {\bibfnamefont {Y.}~\bibnamefont {Zhang}}, \bibinfo {author} {\bibfnamefont {C.~R.}\ \bibnamefont {Kagan}}, \bibinfo {author} {\bibfnamefont {S.~C.}\ \bibnamefont {Glotzer}},\ and\ \bibinfo {author} {\bibfnamefont {C.~B.}\ \bibnamefont {Murray}},\ }\bibfield  {title} {\enquote {\bibinfo {title} {Self-assembly of atomically aligned nanoparticle superlattices from {Pt-Fe$_3$O$_4$} heterodimer nanoparticles},}\ }\href@noop {} {\bibfield  {journal} {\bibinfo  {journal} {Journal of the American Chemical Society}\ }\textbf {\bibinfo {volume} {145}},\ \bibinfo {pages} {6280--6288} (\bibinfo {year} {2023})}\BibitemShut {NoStop}%
\bibitem [{\citenamefont {Marino}\ \emph {et~al.}(2023)\citenamefont {Marino}, \citenamefont {LaCour}, \citenamefont {Moore}, \citenamefont {van Dongen}, \citenamefont {Keller}, \citenamefont {An}, \citenamefont {Yang}, \citenamefont {Rosen}, \citenamefont {Gouget}, \citenamefont {Tsai} \emph {et~al.}}]{marino2023crystallization}%
  \BibitemOpen
  \bibfield  {author} {\bibinfo {author} {\bibfnamefont {E.}~\bibnamefont {Marino}}, \bibinfo {author} {\bibfnamefont {R.~A.}\ \bibnamefont {LaCour}}, \bibinfo {author} {\bibfnamefont {T.~C.}\ \bibnamefont {Moore}}, \bibinfo {author} {\bibfnamefont {S.~W.}\ \bibnamefont {van Dongen}}, \bibinfo {author} {\bibfnamefont {A.~W.}\ \bibnamefont {Keller}}, \bibinfo {author} {\bibfnamefont {D.}~\bibnamefont {An}}, \bibinfo {author} {\bibfnamefont {S.}~\bibnamefont {Yang}}, \bibinfo {author} {\bibfnamefont {D.~J.}\ \bibnamefont {Rosen}}, \bibinfo {author} {\bibfnamefont {G.}~\bibnamefont {Gouget}}, \bibinfo {author} {\bibfnamefont {E.~H.}\ \bibnamefont {Tsai}}, \emph {et~al.},\ }\bibfield  {title} {\enquote {\bibinfo {title} {Crystallization of binary nanocrystal superlattices and the relevance of short-range attraction},}\ }\href@noop {} {\bibfield  {journal} {\bibinfo  {journal} {Nature Synthesis}\ ,\ \bibinfo {pages} {1--12}} (\bibinfo {year} {2023})}\BibitemShut {NoStop}%
\bibitem [{\citenamefont {Ohara}\ \emph {et~al.}(1995)\citenamefont {Ohara}, \citenamefont {Leff}, \citenamefont {Heath},\ and\ \citenamefont {Gelbart}}]{ohara1995crystallization}%
  \BibitemOpen
  \bibfield  {author} {\bibinfo {author} {\bibfnamefont {P.~C.}\ \bibnamefont {Ohara}}, \bibinfo {author} {\bibfnamefont {D.~V.}\ \bibnamefont {Leff}}, \bibinfo {author} {\bibfnamefont {J.~R.}\ \bibnamefont {Heath}},\ and\ \bibinfo {author} {\bibfnamefont {W.~M.}\ \bibnamefont {Gelbart}},\ }\bibfield  {title} {\enquote {\bibinfo {title} {Crystallization of opals from polydisperse nanoparticles},}\ }\href@noop {} {\bibfield  {journal} {\bibinfo  {journal} {Physical Review Letters}\ }\textbf {\bibinfo {volume} {75}},\ \bibinfo {pages} {3466} (\bibinfo {year} {1995})}\BibitemShut {NoStop}%
\bibitem [{\citenamefont {Xia}\ \emph {et~al.}(2011)\citenamefont {Xia}, \citenamefont {Nguyen}, \citenamefont {Yang}, \citenamefont {Lee}, \citenamefont {Santos}, \citenamefont {Podsiadlo}, \citenamefont {Tang}, \citenamefont {Glotzer},\ and\ \citenamefont {Kotov}}]{xia2011self}%
  \BibitemOpen
  \bibfield  {author} {\bibinfo {author} {\bibfnamefont {Y.}~\bibnamefont {Xia}}, \bibinfo {author} {\bibfnamefont {T.~D.}\ \bibnamefont {Nguyen}}, \bibinfo {author} {\bibfnamefont {M.}~\bibnamefont {Yang}}, \bibinfo {author} {\bibfnamefont {B.}~\bibnamefont {Lee}}, \bibinfo {author} {\bibfnamefont {A.}~\bibnamefont {Santos}}, \bibinfo {author} {\bibfnamefont {P.}~\bibnamefont {Podsiadlo}}, \bibinfo {author} {\bibfnamefont {Z.}~\bibnamefont {Tang}}, \bibinfo {author} {\bibfnamefont {S.~C.}\ \bibnamefont {Glotzer}},\ and\ \bibinfo {author} {\bibfnamefont {N.~A.}\ \bibnamefont {Kotov}},\ }\bibfield  {title} {\enquote {\bibinfo {title} {Self-assembly of self-limiting monodisperse supraparticles from polydisperse nanoparticles},}\ }\href@noop {} {\bibfield  {journal} {\bibinfo  {journal} {Nature Nanotechnology}\ }\textbf {\bibinfo {volume} {6}},\ \bibinfo {pages} {580--587} (\bibinfo {year} {2011})}\BibitemShut {NoStop}%
\bibitem [{\citenamefont {Soulantica}\ \emph {et~al.}(2003)\citenamefont {Soulantica}, \citenamefont {Maisonnat}, \citenamefont {Fromen}, \citenamefont {Casanove},\ and\ \citenamefont {Chaudret}}]{soulantica2003spontaneous}%
  \BibitemOpen
  \bibfield  {author} {\bibinfo {author} {\bibfnamefont {K.}~\bibnamefont {Soulantica}}, \bibinfo {author} {\bibfnamefont {A.}~\bibnamefont {Maisonnat}}, \bibinfo {author} {\bibfnamefont {M.-C.}\ \bibnamefont {Fromen}}, \bibinfo {author} {\bibfnamefont {M.-J.}\ \bibnamefont {Casanove}},\ and\ \bibinfo {author} {\bibfnamefont {B.}~\bibnamefont {Chaudret}},\ }\bibfield  {title} {\enquote {\bibinfo {title} {Spontaneous formation of ordered {3D} superlattices of nanocrystals from polydisperse colloidal solutions},}\ }\href@noop {} {\bibfield  {journal} {\bibinfo  {journal} {Angewandte Chemie}\ }\textbf {\bibinfo {volume} {115}},\ \bibinfo {pages} {1989--1993} (\bibinfo {year} {2003})}\BibitemShut {NoStop}%
\bibitem [{\citenamefont {Rabideau}\ \emph {et~al.}(2007)\citenamefont {Rabideau}, \citenamefont {Pell}, \citenamefont {Bonnecaze},\ and\ \citenamefont {Korgel}}]{rabideau2007}%
  \BibitemOpen
  \bibfield  {author} {\bibinfo {author} {\bibfnamefont {B.~D.}\ \bibnamefont {Rabideau}}, \bibinfo {author} {\bibfnamefont {L.~E.}\ \bibnamefont {Pell}}, \bibinfo {author} {\bibfnamefont {R.~T.}\ \bibnamefont {Bonnecaze}},\ and\ \bibinfo {author} {\bibfnamefont {B.~A.}\ \bibnamefont {Korgel}},\ }\bibfield  {title} {\enquote {\bibinfo {title} {Observation of long-range orientational order in monolayers of polydisperse colloids},}\ }\href@noop {} {\bibfield  {journal} {\bibinfo  {journal} {Langmuir}\ }\textbf {\bibinfo {volume} {23}},\ \bibinfo {pages} {1270--1274} (\bibinfo {year} {2007})}\BibitemShut {NoStop}%
\bibitem [{\citenamefont {Sturm}\ and\ \citenamefont {C{\"o}lfen}(2017)}]{sturm2017mesocrystals}%
  \BibitemOpen
  \bibfield  {author} {\bibinfo {author} {\bibfnamefont {E.~V.}\ \bibnamefont {Sturm}}\ and\ \bibinfo {author} {\bibfnamefont {H.}~\bibnamefont {C{\"o}lfen}},\ }\bibfield  {title} {\enquote {\bibinfo {title} {Mesocrystals: {Past}, presence, future},}\ }\href@noop {} {\bibfield  {journal} {\bibinfo  {journal} {Crystals}\ }\textbf {\bibinfo {volume} {7}} (\bibinfo {year} {2017})}\BibitemShut {NoStop}%
\bibitem [{\citenamefont {Jehannin}, \citenamefont {Rao},\ and\ \citenamefont {C{\"o}lfen}(2019)}]{jehannin2019new}%
  \BibitemOpen
  \bibfield  {author} {\bibinfo {author} {\bibfnamefont {M.}~\bibnamefont {Jehannin}}, \bibinfo {author} {\bibfnamefont {A.}~\bibnamefont {Rao}},\ and\ \bibinfo {author} {\bibfnamefont {H.}~\bibnamefont {C{\"o}lfen}},\ }\bibfield  {title} {\enquote {\bibinfo {title} {New horizons of non-classical crystallization},}\ }\href@noop {} {\bibfield  {journal} {\bibinfo  {journal} {Journal of the American Chemical Society}\ } (\bibinfo {year} {2019})}\BibitemShut {NoStop}%
\bibitem [{\citenamefont {Li}, \citenamefont {Senesi},\ and\ \citenamefont {Lee}(2016)}]{li2016small}%
  \BibitemOpen
  \bibfield  {author} {\bibinfo {author} {\bibfnamefont {T.}~\bibnamefont {Li}}, \bibinfo {author} {\bibfnamefont {A.~J.}\ \bibnamefont {Senesi}},\ and\ \bibinfo {author} {\bibfnamefont {B.}~\bibnamefont {Lee}},\ }\bibfield  {title} {\enquote {\bibinfo {title} {Small angle {X-ray} scattering for nanoparticle research},}\ }\href@noop {} {\bibfield  {journal} {\bibinfo  {journal} {Chemical Reviews}\ }\textbf {\bibinfo {volume} {116}},\ \bibinfo {pages} {11128--11180} (\bibinfo {year} {2016})}\BibitemShut {NoStop}%
\bibitem [{\citenamefont {Honecker}\ \emph {et~al.}(2022)\citenamefont {Honecker}, \citenamefont {Bersweiler}, \citenamefont {Erokhin}, \citenamefont {Berkov}, \citenamefont {Chesnel}, \citenamefont {Venero}, \citenamefont {Qdemat}, \citenamefont {Disch}, \citenamefont {Jochum}, \citenamefont {Michels} \emph {et~al.}}]{honecker2022using}%
  \BibitemOpen
  \bibfield  {author} {\bibinfo {author} {\bibfnamefont {D.}~\bibnamefont {Honecker}}, \bibinfo {author} {\bibfnamefont {M.}~\bibnamefont {Bersweiler}}, \bibinfo {author} {\bibfnamefont {S.}~\bibnamefont {Erokhin}}, \bibinfo {author} {\bibfnamefont {D.}~\bibnamefont {Berkov}}, \bibinfo {author} {\bibfnamefont {K.}~\bibnamefont {Chesnel}}, \bibinfo {author} {\bibfnamefont {D.~A.}\ \bibnamefont {Venero}}, \bibinfo {author} {\bibfnamefont {A.}~\bibnamefont {Qdemat}}, \bibinfo {author} {\bibfnamefont {S.}~\bibnamefont {Disch}}, \bibinfo {author} {\bibfnamefont {J.}~\bibnamefont {Jochum}}, \bibinfo {author} {\bibfnamefont {A.}~\bibnamefont {Michels}}, \emph {et~al.},\ }\bibfield  {title} {\enquote {\bibinfo {title} {Using small-angle scattering to guide functional magnetic nanoparticle design},}\ }\href@noop {} {\bibfield  {journal} {\bibinfo  {journal} {Nanoscale Advances}\ } (\bibinfo {year} {2022})}\BibitemShut {NoStop}%
\bibitem [{\citenamefont {Narayanan}\ \emph {et~al.}(2017)\citenamefont {Narayanan}, \citenamefont {Wacklin}, \citenamefont {Konovalov},\ and\ \citenamefont {Lund}}]{narayanan2017recent}%
  \BibitemOpen
  \bibfield  {author} {\bibinfo {author} {\bibfnamefont {T.}~\bibnamefont {Narayanan}}, \bibinfo {author} {\bibfnamefont {H.}~\bibnamefont {Wacklin}}, \bibinfo {author} {\bibfnamefont {O.}~\bibnamefont {Konovalov}},\ and\ \bibinfo {author} {\bibfnamefont {R.}~\bibnamefont {Lund}},\ }\bibfield  {title} {\enquote {\bibinfo {title} {Recent applications of synchrotron radiation and neutrons in the study of soft matter},}\ }\href@noop {} {\bibfield  {journal} {\bibinfo  {journal} {Crystallography Reviews}\ }\textbf {\bibinfo {volume} {23}},\ \bibinfo {pages} {160--226} (\bibinfo {year} {2017})}\BibitemShut {NoStop}%
\bibitem [{\citenamefont {Narayanan}\ and\ \citenamefont {Konovalov}(2020)}]{narayanan2020synchrotron}%
  \BibitemOpen
  \bibfield  {author} {\bibinfo {author} {\bibfnamefont {T.}~\bibnamefont {Narayanan}}\ and\ \bibinfo {author} {\bibfnamefont {O.}~\bibnamefont {Konovalov}},\ }\bibfield  {title} {\enquote {\bibinfo {title} {Synchrotron scattering methods for nanomaterials and soft matter research},}\ }\href@noop {} {\bibfield  {journal} {\bibinfo  {journal} {Materials}\ }\textbf {\bibinfo {volume} {13}},\ \bibinfo {pages} {752} (\bibinfo {year} {2020})}\BibitemShut {NoStop}%
\bibitem [{\citenamefont {Berkovsky}, \citenamefont {Medvedev},\ and\ \citenamefont {Krakov}(1993)}]{berkovsky1993magnetic}%
  \BibitemOpen
  \bibfield  {author} {\bibinfo {author} {\bibfnamefont {B.}~\bibnamefont {Berkovsky}}, \bibinfo {author} {\bibfnamefont {V.~F.}\ \bibnamefont {Medvedev}},\ and\ \bibinfo {author} {\bibfnamefont {M.~S.}\ \bibnamefont {Krakov}},\ }\href@noop {} {\emph {\bibinfo {title} {Magnetic fluids}}}\ (\bibinfo  {publisher} {Oxford Univ. Press},\ \bibinfo {year} {1993})\BibitemShut {NoStop}%
\bibitem [{\citenamefont {Vorobiev}\ \emph {et~al.}(2008)\citenamefont {Vorobiev}, \citenamefont {Chernyshov}, \citenamefont {Gordeev},\ and\ \citenamefont {Orlova}}]{vorobiev2008nondestructive}%
  \BibitemOpen
  \bibfield  {author} {\bibinfo {author} {\bibfnamefont {A.}~\bibnamefont {Vorobiev}}, \bibinfo {author} {\bibfnamefont {D.}~\bibnamefont {Chernyshov}}, \bibinfo {author} {\bibfnamefont {G.}~\bibnamefont {Gordeev}},\ and\ \bibinfo {author} {\bibfnamefont {D.}~\bibnamefont {Orlova}},\ }\bibfield  {title} {\enquote {\bibinfo {title} {Nondestructive characterization of ferrofluids by wide-angle synchrotron light diffraction: crystalline structure and size distribution of colloidal nanoparticles},}\ }\href@noop {} {\bibfield  {journal} {\bibinfo  {journal} {Journal of Applied Crystallography}\ }\textbf {\bibinfo {volume} {41}},\ \bibinfo {pages} {831--835} (\bibinfo {year} {2008})}\BibitemShut {NoStop}%
\bibitem [{\citenamefont {Vorobiev}\ \emph {et~al.}(2015{\natexlab{b}})\citenamefont {Vorobiev}, \citenamefont {Devishvilli}, \citenamefont {Palsson}, \citenamefont {Rundl{\"o}f}, \citenamefont {Johansson}, \citenamefont {Olsson}, \citenamefont {Dennison}, \citenamefont {Wollf}, \citenamefont {Giroud}, \citenamefont {Aguettaz} \emph {et~al.}}]{vorobiev2015recent}%
  \BibitemOpen
  \bibfield  {author} {\bibinfo {author} {\bibfnamefont {A.}~\bibnamefont {Vorobiev}}, \bibinfo {author} {\bibfnamefont {A.}~\bibnamefont {Devishvilli}}, \bibinfo {author} {\bibfnamefont {G.}~\bibnamefont {Palsson}}, \bibinfo {author} {\bibfnamefont {H.}~\bibnamefont {Rundl{\"o}f}}, \bibinfo {author} {\bibfnamefont {N.}~\bibnamefont {Johansson}}, \bibinfo {author} {\bibfnamefont {A.}~\bibnamefont {Olsson}}, \bibinfo {author} {\bibfnamefont {A.}~\bibnamefont {Dennison}}, \bibinfo {author} {\bibfnamefont {M.}~\bibnamefont {Wollf}}, \bibinfo {author} {\bibfnamefont {B.}~\bibnamefont {Giroud}}, \bibinfo {author} {\bibfnamefont {O.}~\bibnamefont {Aguettaz}}, \emph {et~al.},\ }\bibfield  {title} {\enquote {\bibinfo {title} {Recent upgrade of the polarized neutron reflectometer super adam},}\ }\href@noop {} {\bibfield  {journal} {\bibinfo  {journal} {Neutron News}\ }\textbf {\bibinfo {volume} {26}},\ \bibinfo {pages} {25--26} (\bibinfo {year} {2015}{\natexlab{b}})}\BibitemShut {NoStop}%
\bibitem [{\citenamefont {Harkins}(1952)}]{20}%
  \BibitemOpen
  \bibfield  {author} {\bibinfo {author} {\bibfnamefont {W.~D.}\ \bibnamefont {Harkins}},\ }\href@noop {} {\emph {\bibinfo {title} {The physical chemistry of surface films}}}\ (\bibinfo  {publisher} {Reinhold},\ \bibinfo {year} {1952})\BibitemShut {NoStop}%
\bibitem [{\citenamefont {Mishra}\ \emph {et~al.}(2012)\citenamefont {Mishra}, \citenamefont {Benitez}, \citenamefont {Petracic}, \citenamefont {Confalonieri}, \citenamefont {Szary}, \citenamefont {Br{\"u}ssing}, \citenamefont {Theis-Br{\"o}hl}, \citenamefont {Devishvili}, \citenamefont {Vorobiev}, \citenamefont {Konovalov} \emph {et~al.}}]{mishra2012self}%
  \BibitemOpen
  \bibfield  {author} {\bibinfo {author} {\bibfnamefont {D.}~\bibnamefont {Mishra}}, \bibinfo {author} {\bibfnamefont {M.}~\bibnamefont {Benitez}}, \bibinfo {author} {\bibfnamefont {O.}~\bibnamefont {Petracic}}, \bibinfo {author} {\bibfnamefont {G.~B.}\ \bibnamefont {Confalonieri}}, \bibinfo {author} {\bibfnamefont {P.}~\bibnamefont {Szary}}, \bibinfo {author} {\bibfnamefont {F.}~\bibnamefont {Br{\"u}ssing}}, \bibinfo {author} {\bibfnamefont {K.}~\bibnamefont {Theis-Br{\"o}hl}}, \bibinfo {author} {\bibfnamefont {A.}~\bibnamefont {Devishvili}}, \bibinfo {author} {\bibfnamefont {A.}~\bibnamefont {Vorobiev}}, \bibinfo {author} {\bibfnamefont {O.}~\bibnamefont {Konovalov}}, \emph {et~al.},\ }\bibfield  {title} {\enquote {\bibinfo {title} {Self-assembled iron oxide nanoparticle multilayer: x-ray and polarized neutron reflectivity},}\ }\href@noop {} {\bibfield  {journal} {\bibinfo  {journal} {Nanotechnology}\ }\textbf {\bibinfo {volume} {23}},\ \bibinfo {pages} {055707} (\bibinfo {year} {2012})}\BibitemShut
  {NoStop}%
\bibitem [{\citenamefont {Theis-Br{\"o}ohl}\ \emph {et~al.}(2018)\citenamefont {Theis-Br{\"o}ohl}, \citenamefont {Vreeland}, \citenamefont {Gomez}, \citenamefont {Huber}, \citenamefont {Saini}, \citenamefont {Wolff}, \citenamefont {Maranville}, \citenamefont {Brok}, \citenamefont {Krycka}, \citenamefont {Dura} \emph {et~al.}}]{theis2018self}%
  \BibitemOpen
  \bibfield  {author} {\bibinfo {author} {\bibfnamefont {K.}~\bibnamefont {Theis-Br{\"o}ohl}}, \bibinfo {author} {\bibfnamefont {E.~C.}\ \bibnamefont {Vreeland}}, \bibinfo {author} {\bibfnamefont {A.}~\bibnamefont {Gomez}}, \bibinfo {author} {\bibfnamefont {D.~L.}\ \bibnamefont {Huber}}, \bibinfo {author} {\bibfnamefont {A.}~\bibnamefont {Saini}}, \bibinfo {author} {\bibfnamefont {M.}~\bibnamefont {Wolff}}, \bibinfo {author} {\bibfnamefont {B.~B.}\ \bibnamefont {Maranville}}, \bibinfo {author} {\bibfnamefont {E.}~\bibnamefont {Brok}}, \bibinfo {author} {\bibfnamefont {K.~L.}\ \bibnamefont {Krycka}}, \bibinfo {author} {\bibfnamefont {J.~A.}\ \bibnamefont {Dura}}, \emph {et~al.},\ }\bibfield  {title} {\enquote {\bibinfo {title} {Self-assembled layering of magnetic nanoparticles in a ferrofluid on silicon surfaces},}\ }\href@noop {} {\bibfield  {journal} {\bibinfo  {journal} {ACS applied materials \& interfaces}\ }\textbf {\bibinfo {volume} {10}},\ \bibinfo {pages} {5050--5060} (\bibinfo {year}
  {2018})}\BibitemShut {NoStop}%
\bibitem [{\citenamefont {Bender}\ \emph {et~al.}(2018)\citenamefont {Bender}, \citenamefont {Wetterskog}, \citenamefont {Honecker}, \citenamefont {Fock}, \citenamefont {Frandsen}, \citenamefont {Moerland}, \citenamefont {Bogart}, \citenamefont {Posth}, \citenamefont {Szczerba}, \citenamefont {Gavil{\'a}n} \emph {et~al.}}]{bender2018dipolar}%
  \BibitemOpen
  \bibfield  {author} {\bibinfo {author} {\bibfnamefont {P.}~\bibnamefont {Bender}}, \bibinfo {author} {\bibfnamefont {E.}~\bibnamefont {Wetterskog}}, \bibinfo {author} {\bibfnamefont {D.}~\bibnamefont {Honecker}}, \bibinfo {author} {\bibfnamefont {J.}~\bibnamefont {Fock}}, \bibinfo {author} {\bibfnamefont {C.}~\bibnamefont {Frandsen}}, \bibinfo {author} {\bibfnamefont {C.}~\bibnamefont {Moerland}}, \bibinfo {author} {\bibfnamefont {L.~K.}\ \bibnamefont {Bogart}}, \bibinfo {author} {\bibfnamefont {O.}~\bibnamefont {Posth}}, \bibinfo {author} {\bibfnamefont {W.}~\bibnamefont {Szczerba}}, \bibinfo {author} {\bibfnamefont {H.}~\bibnamefont {Gavil{\'a}n}}, \emph {et~al.},\ }\bibfield  {title} {\enquote {\bibinfo {title} {Dipolar-coupled moment correlations in clusters of magnetic nanoparticles},}\ }\href@noop {} {\bibfield  {journal} {\bibinfo  {journal} {Physical Review B}\ }\textbf {\bibinfo {volume} {98}},\ \bibinfo {pages} {224420} (\bibinfo {year} {2018})}\BibitemShut {NoStop}%
\bibitem [{\citenamefont {Theis-Br{\"o}hl}\ \emph {et~al.}(2020)\citenamefont {Theis-Br{\"o}hl}, \citenamefont {Saini}, \citenamefont {Wolff}, \citenamefont {Dura}, \citenamefont {Maranville},\ and\ \citenamefont {Borchers}}]{theis2020self}%
  \BibitemOpen
  \bibfield  {author} {\bibinfo {author} {\bibfnamefont {K.}~\bibnamefont {Theis-Br{\"o}hl}}, \bibinfo {author} {\bibfnamefont {A.}~\bibnamefont {Saini}}, \bibinfo {author} {\bibfnamefont {M.}~\bibnamefont {Wolff}}, \bibinfo {author} {\bibfnamefont {J.~A.}\ \bibnamefont {Dura}}, \bibinfo {author} {\bibfnamefont {B.~B.}\ \bibnamefont {Maranville}},\ and\ \bibinfo {author} {\bibfnamefont {J.~A.}\ \bibnamefont {Borchers}},\ }\bibfield  {title} {\enquote {\bibinfo {title} {Self-assembly of magnetic nanoparticles in ferrofluids on different templates investigated by neutron reflectometry},}\ }\href@noop {} {\bibfield  {journal} {\bibinfo  {journal} {Nanomaterials}\ }\textbf {\bibinfo {volume} {10}},\ \bibinfo {pages} {1231} (\bibinfo {year} {2020})}\BibitemShut {NoStop}%
\bibitem [{\citenamefont {Chesnel}\ \emph {et~al.}(2018)\citenamefont {Chesnel}, \citenamefont {Griner}, \citenamefont {Smith}, \citenamefont {Cai}, \citenamefont {Trevino}, \citenamefont {Newbold}, \citenamefont {Wang}, \citenamefont {Liu}, \citenamefont {Jal}, \citenamefont {Reid} \emph {et~al.}}]{chesnel2018unraveling}%
  \BibitemOpen
  \bibfield  {author} {\bibinfo {author} {\bibfnamefont {K.}~\bibnamefont {Chesnel}}, \bibinfo {author} {\bibfnamefont {D.}~\bibnamefont {Griner}}, \bibinfo {author} {\bibfnamefont {D.}~\bibnamefont {Smith}}, \bibinfo {author} {\bibfnamefont {Y.}~\bibnamefont {Cai}}, \bibinfo {author} {\bibfnamefont {M.}~\bibnamefont {Trevino}}, \bibinfo {author} {\bibfnamefont {B.}~\bibnamefont {Newbold}}, \bibinfo {author} {\bibfnamefont {T.}~\bibnamefont {Wang}}, \bibinfo {author} {\bibfnamefont {T.}~\bibnamefont {Liu}}, \bibinfo {author} {\bibfnamefont {E.}~\bibnamefont {Jal}}, \bibinfo {author} {\bibfnamefont {A.}~\bibnamefont {Reid}}, \emph {et~al.},\ }\bibfield  {title} {\enquote {\bibinfo {title} {Unraveling nanoscale magnetic ordering in {Fe$_3$O$_4$} nanoparticle assemblies via {X-rays}},}\ }\href@noop {} {\bibfield  {journal} {\bibinfo  {journal} {Magnetochemistry}\ }\textbf {\bibinfo {volume} {4}},\ \bibinfo {pages} {42} (\bibinfo {year} {2018})}\BibitemShut {NoStop}%
\bibitem [{\citenamefont {Wang}\ \emph {et~al.}(2019)\citenamefont {Wang}, \citenamefont {Qdemat}, \citenamefont {Petracic}, \citenamefont {Kentzinger}, \citenamefont {R{\"u}cker}, \citenamefont {Zheng}, \citenamefont {Lu}, \citenamefont {Wei}, \citenamefont {Dunin-Borkowski},\ and\ \citenamefont {Br{\"u}ckel}}]{wang2019manipulation}%
  \BibitemOpen
  \bibfield  {author} {\bibinfo {author} {\bibfnamefont {L.-M.}\ \bibnamefont {Wang}}, \bibinfo {author} {\bibfnamefont {A.}~\bibnamefont {Qdemat}}, \bibinfo {author} {\bibfnamefont {O.}~\bibnamefont {Petracic}}, \bibinfo {author} {\bibfnamefont {E.}~\bibnamefont {Kentzinger}}, \bibinfo {author} {\bibfnamefont {U.}~\bibnamefont {R{\"u}cker}}, \bibinfo {author} {\bibfnamefont {F.}~\bibnamefont {Zheng}}, \bibinfo {author} {\bibfnamefont {P.-H.}\ \bibnamefont {Lu}}, \bibinfo {author} {\bibfnamefont {X.-K.}\ \bibnamefont {Wei}}, \bibinfo {author} {\bibfnamefont {R.~E.}\ \bibnamefont {Dunin-Borkowski}},\ and\ \bibinfo {author} {\bibfnamefont {T.}~\bibnamefont {Br{\"u}ckel}},\ }\bibfield  {title} {\enquote {\bibinfo {title} {Manipulation of dipolar magnetism in low-dimensional iron oxide nanoparticle assemblies},}\ }\href@noop {} {\bibfield  {journal} {\bibinfo  {journal} {Physical Chemistry Chemical Physics}\ }\textbf {\bibinfo {volume} {21}},\ \bibinfo {pages} {6171--6177} (\bibinfo {year} {2019})}\BibitemShut
  {NoStop}%
\bibitem [{\citenamefont {Ukleev}, \citenamefont {Snigireva},\ and\ \citenamefont {Vorobiev}(2017)}]{ukleev2017polarized}%
  \BibitemOpen
  \bibfield  {author} {\bibinfo {author} {\bibfnamefont {V.}~\bibnamefont {Ukleev}}, \bibinfo {author} {\bibfnamefont {I.}~\bibnamefont {Snigireva}},\ and\ \bibinfo {author} {\bibfnamefont {A.}~\bibnamefont {Vorobiev}},\ }\bibfield  {title} {\enquote {\bibinfo {title} {Polarized neutron reflectometry study from iron oxide nanoparticles monolayer},}\ }\href@noop {} {\bibfield  {journal} {\bibinfo  {journal} {Surfaces and Interfaces}\ }\textbf {\bibinfo {volume} {9}},\ \bibinfo {pages} {143--146} (\bibinfo {year} {2017})}\BibitemShut {NoStop}%
\bibitem [{\citenamefont {Zhu}(2005)}]{zhu2005modern}%
  \BibitemOpen
  \bibfield  {author} {\bibinfo {author} {\bibfnamefont {Y.}~\bibnamefont {Zhu}},\ }\href@noop {} {\emph {\bibinfo {title} {Modern techniques for characterizing magnetic materials}}}\ (\bibinfo  {publisher} {Springer Science \& Business Media},\ \bibinfo {year} {2005})\BibitemShut {NoStop}%
\bibitem [{\citenamefont {Bj{\"o}rck}\ and\ \citenamefont {Andersson}(2007)}]{bjorck2007genx}%
  \BibitemOpen
  \bibfield  {author} {\bibinfo {author} {\bibfnamefont {M.}~\bibnamefont {Bj{\"o}rck}}\ and\ \bibinfo {author} {\bibfnamefont {G.}~\bibnamefont {Andersson}},\ }\bibfield  {title} {\enquote {\bibinfo {title} {Genx: an extensible {X-ray} reflectivity refinement program utilizing differential evolution},}\ }\href@noop {} {\bibfield  {journal} {\bibinfo  {journal} {Journal of Applied Crystallography}\ }\textbf {\bibinfo {volume} {40}},\ \bibinfo {pages} {1174--1178} (\bibinfo {year} {2007})}\BibitemShut {NoStop}%
\bibitem [{\citenamefont {Glavic}\ and\ \citenamefont {Bj{\"o}rck}(2022)}]{glavic2022genx}%
  \BibitemOpen
  \bibfield  {author} {\bibinfo {author} {\bibfnamefont {A.}~\bibnamefont {Glavic}}\ and\ \bibinfo {author} {\bibfnamefont {M.}~\bibnamefont {Bj{\"o}rck}},\ }\bibfield  {title} {\enquote {\bibinfo {title} {{GenX 3}: the latest generation of an established tool},}\ }\href@noop {} {\bibfield  {journal} {\bibinfo  {journal} {Journal of applied crystallography}\ }\textbf {\bibinfo {volume} {55}} (\bibinfo {year} {2022})}\BibitemShut {NoStop}%
\bibitem [{\citenamefont {Pichon}\ \emph {et~al.}(2014)\citenamefont {Pichon}, \citenamefont {Leuvey}, \citenamefont {Ihawakrim}, \citenamefont {Bernard}, \citenamefont {Schmerber},\ and\ \citenamefont {Begin-Colin}}]{pichon2014magnetic}%
  \BibitemOpen
  \bibfield  {author} {\bibinfo {author} {\bibfnamefont {B.~P.}\ \bibnamefont {Pichon}}, \bibinfo {author} {\bibfnamefont {C.}~\bibnamefont {Leuvey}}, \bibinfo {author} {\bibfnamefont {D.}~\bibnamefont {Ihawakrim}}, \bibinfo {author} {\bibfnamefont {P.}~\bibnamefont {Bernard}}, \bibinfo {author} {\bibfnamefont {G.}~\bibnamefont {Schmerber}},\ and\ \bibinfo {author} {\bibfnamefont {S.}~\bibnamefont {Begin-Colin}},\ }\bibfield  {title} {\enquote {\bibinfo {title} {Magnetic properties of mono-and multilayer assemblies of iron oxide nanoparticles promoted by {SAMs}},}\ }\href@noop {} {\bibfield  {journal} {\bibinfo  {journal} {The Journal of Physical Chemistry C}\ }\textbf {\bibinfo {volume} {118}},\ \bibinfo {pages} {3828--3837} (\bibinfo {year} {2014})}\BibitemShut {NoStop}%
\bibitem [{\citenamefont {Ukleev}\ \emph {et~al.}(2016)\citenamefont {Ukleev}, \citenamefont {Khassanov}, \citenamefont {Snigireva}, \citenamefont {Konovalov},\ and\ \citenamefont {Vorobiev}}]{ukleev2016x}%
  \BibitemOpen
  \bibfield  {author} {\bibinfo {author} {\bibfnamefont {V.}~\bibnamefont {Ukleev}}, \bibinfo {author} {\bibfnamefont {A.}~\bibnamefont {Khassanov}}, \bibinfo {author} {\bibfnamefont {I.}~\bibnamefont {Snigireva}}, \bibinfo {author} {\bibfnamefont {O.}~\bibnamefont {Konovalov}},\ and\ \bibinfo {author} {\bibfnamefont {A.}~\bibnamefont {Vorobiev}},\ }\bibfield  {title} {\enquote {\bibinfo {title} {X-ray scattering characterization of iron oxide nanoparticles {Langmuir} film on water surface and on a solid substrate},}\ }\href@noop {} {\bibfield  {journal} {\bibinfo  {journal} {Thin Solid Films}\ }\textbf {\bibinfo {volume} {616}},\ \bibinfo {pages} {43--47} (\bibinfo {year} {2016})}\BibitemShut {NoStop}%
\bibitem [{\citenamefont {Barbeta}\ \emph {et~al.}(2010)\citenamefont {Barbeta}, \citenamefont {Jardim}, \citenamefont {Kiyohara}, \citenamefont {Effenberger},\ and\ \citenamefont {Rossi}}]{barbeta2010magnetic}%
  \BibitemOpen
  \bibfield  {author} {\bibinfo {author} {\bibfnamefont {V.}~\bibnamefont {Barbeta}}, \bibinfo {author} {\bibfnamefont {R.}~\bibnamefont {Jardim}}, \bibinfo {author} {\bibfnamefont {P.}~\bibnamefont {Kiyohara}}, \bibinfo {author} {\bibfnamefont {F.}~\bibnamefont {Effenberger}},\ and\ \bibinfo {author} {\bibfnamefont {L.}~\bibnamefont {Rossi}},\ }\bibfield  {title} {\enquote {\bibinfo {title} {Magnetic properties of {Fe$_3$O$_4$} nanoparticles coated with oleic and dodecanoic acids},}\ }\href@noop {} {\bibfield  {journal} {\bibinfo  {journal} {Journal of Applied Physics}\ }\textbf {\bibinfo {volume} {107}},\ \bibinfo {pages} {073913} (\bibinfo {year} {2010})}\BibitemShut {NoStop}%
\bibitem [{\citenamefont {Frandsen}\ \emph {et~al.}(2021)\citenamefont {Frandsen}, \citenamefont {Read}, \citenamefont {Stevens}, \citenamefont {Walker}, \citenamefont {Christiansen}, \citenamefont {Harrison},\ and\ \citenamefont {Chesnel}}]{frandsen2021superparamagnetic}%
  \BibitemOpen
  \bibfield  {author} {\bibinfo {author} {\bibfnamefont {B.~A.}\ \bibnamefont {Frandsen}}, \bibinfo {author} {\bibfnamefont {C.}~\bibnamefont {Read}}, \bibinfo {author} {\bibfnamefont {J.}~\bibnamefont {Stevens}}, \bibinfo {author} {\bibfnamefont {C.}~\bibnamefont {Walker}}, \bibinfo {author} {\bibfnamefont {M.}~\bibnamefont {Christiansen}}, \bibinfo {author} {\bibfnamefont {R.~G.}\ \bibnamefont {Harrison}},\ and\ \bibinfo {author} {\bibfnamefont {K.}~\bibnamefont {Chesnel}},\ }\bibfield  {title} {\enquote {\bibinfo {title} {Superparamagnetic dynamics and blocking transition in {Fe$_3$O$_4$} nanoparticles probed by vibrating sample magnetometry and muon spin relaxation},}\ }\href@noop {} {\bibfield  {journal} {\bibinfo  {journal} {Physical Review Materials}\ }\textbf {\bibinfo {volume} {5}},\ \bibinfo {pages} {054411} (\bibinfo {year} {2021})}\BibitemShut {NoStop}%
\bibitem [{\citenamefont {Rackham}\ \emph {et~al.}(2023)\citenamefont {Rackham}, \citenamefont {Pratt}, \citenamefont {Griner}, \citenamefont {Smith}, \citenamefont {Cai}, \citenamefont {Harrison}, \citenamefont {Reid}, \citenamefont {Kortright}, \citenamefont {Transtrum},\ and\ \citenamefont {Chesnel}}]{rackham2023field}%
  \BibitemOpen
  \bibfield  {author} {\bibinfo {author} {\bibfnamefont {J.}~\bibnamefont {Rackham}}, \bibinfo {author} {\bibfnamefont {B.}~\bibnamefont {Pratt}}, \bibinfo {author} {\bibfnamefont {D.}~\bibnamefont {Griner}}, \bibinfo {author} {\bibfnamefont {D.}~\bibnamefont {Smith}}, \bibinfo {author} {\bibfnamefont {Y.}~\bibnamefont {Cai}}, \bibinfo {author} {\bibfnamefont {R.~G.}\ \bibnamefont {Harrison}}, \bibinfo {author} {\bibfnamefont {A.}~\bibnamefont {Reid}}, \bibinfo {author} {\bibfnamefont {J.}~\bibnamefont {Kortright}}, \bibinfo {author} {\bibfnamefont {M.~K.}\ \bibnamefont {Transtrum}},\ and\ \bibinfo {author} {\bibfnamefont {K.}~\bibnamefont {Chesnel}},\ }\bibfield  {title} {\enquote {\bibinfo {title} {Field-dependent nanospin ordering in monolayers of {Fe$_3$O$_4$} nanoparticles throughout the superparamagnetic blocking transition},}\ }\href@noop {} {\bibfield  {journal} {\bibinfo  {journal} {Physical Review B}\ }\textbf {\bibinfo {volume} {108}},\ \bibinfo {pages} {104415} (\bibinfo {year}
  {2023})}\BibitemShut {NoStop}%
\bibitem [{\citenamefont {Sahoo}\ \emph {et~al.}(2004)\citenamefont {Sahoo}, \citenamefont {Cheon}, \citenamefont {Wang}, \citenamefont {Luo}, \citenamefont {Furlani},\ and\ \citenamefont {Prasad}}]{sahoo2004field}%
  \BibitemOpen
  \bibfield  {author} {\bibinfo {author} {\bibfnamefont {Y.}~\bibnamefont {Sahoo}}, \bibinfo {author} {\bibfnamefont {M.}~\bibnamefont {Cheon}}, \bibinfo {author} {\bibfnamefont {S.}~\bibnamefont {Wang}}, \bibinfo {author} {\bibfnamefont {H.}~\bibnamefont {Luo}}, \bibinfo {author} {\bibfnamefont {E.}~\bibnamefont {Furlani}},\ and\ \bibinfo {author} {\bibfnamefont {P.}~\bibnamefont {Prasad}},\ }\bibfield  {title} {\enquote {\bibinfo {title} {Field-directed self-assembly of magnetic nanoparticles},}\ }\href@noop {} {\bibfield  {journal} {\bibinfo  {journal} {The Journal of Physical Chemistry B}\ }\textbf {\bibinfo {volume} {108}},\ \bibinfo {pages} {3380--3383} (\bibinfo {year} {2004})}\BibitemShut {NoStop}%
\bibitem [{\citenamefont {Tracy}\ and\ \citenamefont {Crawford}(2013)}]{tracy2013magnetic}%
  \BibitemOpen
  \bibfield  {author} {\bibinfo {author} {\bibfnamefont {J.~B.}\ \bibnamefont {Tracy}}\ and\ \bibinfo {author} {\bibfnamefont {T.~M.}\ \bibnamefont {Crawford}},\ }\bibfield  {title} {\enquote {\bibinfo {title} {Magnetic field-directed self-assembly of magnetic nanoparticles},}\ }\href@noop {} {\bibfield  {journal} {\bibinfo  {journal} {MRS bulletin}\ }\textbf {\bibinfo {volume} {38}},\ \bibinfo {pages} {915--920} (\bibinfo {year} {2013})}\BibitemShut {NoStop}%
\bibitem [{\citenamefont {Mehdizadeh~Taheri}\ \emph {et~al.}(2015)\citenamefont {Mehdizadeh~Taheri}, \citenamefont {Michaelis}, \citenamefont {Friedrich}, \citenamefont {F{\"o}rster}, \citenamefont {Drechsler}, \citenamefont {R{\"o}mer}, \citenamefont {B{\"o}secke}, \citenamefont {Narayanan}, \citenamefont {Weber}, \citenamefont {Rehberg} \emph {et~al.}}]{mehdizadeh2015self}%
  \BibitemOpen
  \bibfield  {author} {\bibinfo {author} {\bibfnamefont {S.}~\bibnamefont {Mehdizadeh~Taheri}}, \bibinfo {author} {\bibfnamefont {M.}~\bibnamefont {Michaelis}}, \bibinfo {author} {\bibfnamefont {T.}~\bibnamefont {Friedrich}}, \bibinfo {author} {\bibfnamefont {B.}~\bibnamefont {F{\"o}rster}}, \bibinfo {author} {\bibfnamefont {M.}~\bibnamefont {Drechsler}}, \bibinfo {author} {\bibfnamefont {F.~M.}\ \bibnamefont {R{\"o}mer}}, \bibinfo {author} {\bibfnamefont {P.}~\bibnamefont {B{\"o}secke}}, \bibinfo {author} {\bibfnamefont {T.}~\bibnamefont {Narayanan}}, \bibinfo {author} {\bibfnamefont {B.}~\bibnamefont {Weber}}, \bibinfo {author} {\bibfnamefont {I.}~\bibnamefont {Rehberg}}, \emph {et~al.},\ }\bibfield  {title} {\enquote {\bibinfo {title} {Self-assembly of smallest magnetic particles},}\ }\href@noop {} {\bibfield  {journal} {\bibinfo  {journal} {Proceedings of the National Academy of Sciences}\ }\textbf {\bibinfo {volume} {112}},\ \bibinfo {pages} {14484--14489} (\bibinfo {year} {2015})}\BibitemShut {NoStop}%
\end{thebibliography}%
\end{document}